\documentclass[hidelinks,onefignum,onetabnum]{siamart250211}

\usepackage{lipsum}
\usepackage{amsfonts}
\usepackage{graphicx}
\usepackage{epstopdf}
\usepackage{algorithmic}

\ifpdf
  \DeclareGraphicsExtensions{.eps,.pdf,.png,.jpg}
\else
  \DeclareGraphicsExtensions{.eps}
\fi

\newsiamremark{remark}{Remark}
\newsiamremark{hypothesis}{Hypothesis}
\crefname{hypothesis}{Hypothesis}{Hypotheses}
\newsiamthm{claim}{Claim}
\newsiamremark{fact}{Fact}
\crefname{fact}{Fact}{Facts}

\headers{}{K. Shaju, T. Laepple, and P. Zaspel}

\title{Bayesian inference: Kernel-based model for surface temperature reconstruction in ice borehole thermometry}

\author{
Kshema Shaju\footnotemark[3]
\thanks{
Alfred Wegener Institute, Helmholtz Centre for Polar and Marine Research, Potsdam, Germany}
\and
Thomas Laepple\footnotemark[1]
    \thanks{Department of Geosciences, University of Bremen, and MARUM, University of Bremen, Germany}
\and
Peter Zaspel\thanks{School of Mathematics and Natural Sciences, Bergische Universität Wuppertal, Germany 
(\email{zaspel@uni-wuppertal.de})}
}

\usepackage{amsopn}

\begin{document}

\maketitle

\begin{abstract}
Reconstructing past surface temperature from shallow ice borehole temperature profiles requires solving an ill-posed inverse problem while quantifying uncertainties arising from measurements and prior assumptions. Bayesian formulations enable probabilistic reconstruction of surface temperature histories and uncertainty quantification. Existing reversible jump-Markov chain Monte Carlo approach based on adaptive piecewise-linear surface temperature models can, however, be computationally demanding. Here, we introduce a kernel-based surface temperature model that enables the use of a parallel ensemble Markov chain Monte Carlo sampler for efficient exploration of the solution space and quantification of the posterior. Using synthetic experiments, we investigate the effects of kernel configuration, measurement uncertainty, measurement density, and temporal smearing on reconstruction performance. We find that reconstruction quality is largely insensitive to the number of kernels once the kernel basis is sufficiently dense. Reducing measurement uncertainty substantially improves reconstructions, whereas increasing the number of borehole temperature measurements provides only marginal benefit. Finally, we evaluate the method using realistic surrogate climate histories that combine long-term temperature changes with stochastic climate variability. The kernel-based surface temperature model cannot represent short-term variability and therefore cannot fully explain the realistic measurements, highlighting the need to account for this approximation uncertainty. The likelihood is adapted to include the approximation uncertainty of the surface temperature model, yielding robust reconstructions with reliable posterior uncertainties. Overall, our results demonstrate that kernel-based Bayesian inversion provides an efficient framework for shallow ice borehole based climate reconstructions.

\end{abstract}



\section{Introduction}

Polar ice sheet represents the greatest potential source of global sea-level rise, and its response to climate change is a key source of uncertainty for future projections. Instrumental surface temperature records in the polar regions are generally too short to characterize long-term climate variability, making climate proxy archives essential for reconstructing past climate. Climate proxies are natural archives of past climate information, ice boreholes provide one such archive. Variations in surface temperature propagate into the ice by heat diffusion and advection, leaving a subsurface temperature profile that contains information about past surface temperature evolution.

The forward (direct) problem, the propagation of a given surface temperature history into the ice, is described by a heat transfer model, which simulates  the resulting temperature distribution with depth and time. To infer the surface temperature evolution that caused the present day heat distribution in ice that is observed via ice borehole temperature measurements, we need to solve an inverse problem-- as we are trying to determine the cause, the surface temperature variations, from the effect, the heat distribution in ice. Various approaches have been employed to address this ill-posed inverse problem of surface temperature reconstructions using borehole thermometry \cite{Kakuta1992, MareschalBeltrami1992, Shen1992svd, DahlJensen1998,WoodburyFerguson2006,Hopcroft2007inference, Muto2011recent}; however, most of these approaches impose constraints on the smoothness or resolution of the reconstructed surface temperature history \cite{Hopcroft2007inference}. Formulating the problem in a Bayesian framework allows us to reconstruct surface temperature and quantify uncertainty by incorporating measurement error and prior belief specified in a probabilistic way \cite{Hopcroft2007inference, muto2011multidecadal, Muto2011recent, reich2015probabilistic, Hopcroft2023global, Groenke2023thermalstate}. 

In developing a Bayesian inverse model for surface temperature reconstruction, the representation of surface temperature history plays a central role. Parametrization of surface temperature should be done in a way that reconstructions have minimal dependency on the particular choice of parameters used in the surface temperature model. Bayesian inference solves the inverse problem by quantifying the posterior uncertainty of these model parameters by updating their respective prior probabilities based on measured data. Therefore, posterior probability represents the uncertainty of model parameters after taking measurements \cite{reich2015probabilistic}.
 
Posterior distributions of parameters can be approximated using Markov chain Monte Carlo methods (MCMC) \cite{Hastings1970, Metropolis1953}. Surface temperature reconstructions from land and ice boreholes were performed using a Bayesian approach \cite{Hopcroft2007inference, muto2011multidecadal} employing reversible jump Markov chain Monte Carlo (RJ-MCMC) sampling using an adaptive piecewise-linear surface temperature model. While this approach ensures that the reconstructions are not dependent on parameters, still it employs a single chain serial process, making it difficult to efficiently explore all possible time series with respect to the given prior information on surface temperature evolution.   

Surface temperature reconstruction from shallow ice boreholes is particularly challenging because only the upper few hundred meters of borehole measurements are available, whereas the governing heat transport acts over the full ice thickness of approximately $2.5-3 ~\mathrm{km}$. Consequently, the available observations contain only limited information about past surface temperatures, resulting in a large space of plausible solution. In other words, the solution space of surface temperatures will be very large with such limited data \cite{bishop2006prml}. Also, the multiple solutions that fit the measurements can be quite wide. Capturing this uncertainty requires sufficiently flexible surface temperature models without unnecessarily restricting the solution space. However, increased model flexibility increases the number of model parameters and hence the dimensionality of the inverse problem, making efficient exploration of the parameter space increasingly challenging.  The lack of proper exploration of solution space lead to unreliable posterior summaries and misleading uncertainty assessment \cite{roy2020convergence}. This motivates the use of efficient parallel ensemble MCMC methods for posterior sampling.

In this study, we introduce a kernel-based representation of surface temperature history that enables the use of a parallel ensemble MCMC sampler for efficient posterior exploration. We first evaluate the proposed approach using synthetic surface temperature histories to investigate the influence of kernel configuration, measurement uncertainty, measurement density, and temporal smearing on reconstruction performance. We then compare the proposed method with the adaptive piecewise-linear RJ-MCMC approach. Finally, we assess its performance using realistic surrogate climate histories that combine long-term temperature trends with stochastic climate variability, and diagnose the influence of unresolved short-term climate variability in measurements on the reconstructions. We further demonstrate a strategy for obtaining reliable uncertainty estimates and robust surface temperature reconstructions despite unresolved fast climate variations.

\section{Inverse problem of climate reconstruction from ice-borehole}
\label{sec:inv_prob}

We consider the inverse problem of determining past surface temperature from observed data of ice-borehole temperature measurements. The surface temperature $\theta(t)$ is a time-dependent parameter, assumed to vary continuously from $t_0$ years in the past to the present time $t_{\mathrm{pr}}$. The observed data is modeled deterministically as,
\begin{equation}
\label{eqn_dm_noise}
     \vec{y} = \tilde{T}(\theta(t)) + \vec{\epsilon}
     \end{equation}
where $\tilde{T}$ is the forward operator that describes the governing equations mapping surface temperature $\theta(t)$ to the observed borehole data $\vec{y}$. $\vec{\epsilon}$ is the error in observing data (measurement error). $\vec{y}$ and $\vec{\epsilon}$ are $n$-dimensional real vectors, and $n$ is the number of temperature measurements. Here, the governing equation is the heat diffusion advection equation that describes the mapping of past surface temperature to the present day borehole temperature data. The inverse problem is hence defined as the recovery of the unknown surface temperature $\theta(t) \in \mathbb{X}$ from the imperfect observation $\vec{y} \in \mathbb{Y}$ of $\tilde{T}(\theta(t))$. Here, $\tilde{T} : \mathbb{X} \rightarrow \mathbb{Y}$  denotes the forward operator, $\mathbb{X}$ is the (potentially infinite-dimensional) space of admissible surface temperature histories, and $\mathbb{Y}$ is the finite-dimensional observation space \cite{lie2018random} with each of its realizations being an 
$n$-dimensional real vector (for example $\vec{y}$). 

The inverse problem is ill-posed as there may be no element $\theta(t) \in \mathbb{X}$ for which $\tilde{T}(\theta(t))$ gives $\vec{y}$, or there may be several $\theta(t) \in \mathbb{X}$ for which $\tilde{T}(\theta(t))$ gives $\vec{y}$ or $\theta(t)$ is very sensitive to the observed data $\vec{y}$ \cite{lie2018random, scheichl2013large}.

\subsection{Forward model}
\label{subsec:fwd_model}
The forward operator $\tilde{T}$ maps a surface temperature history $\theta(t)$ to the borehole temperature measurements. Determining the full continuous temperature profile is done by solving the one-dimensional heat diffusion–advection equation,
\begin{equation}\label{eq_hae}
\frac{\partial T_\theta}{\partial t}= k\frac{\partial^2 T_\theta}{\partial z^2}-w\frac{\partial T_\theta}{\partial z}\,,
\end{equation}
where $T_\theta$ is the temperature as a function of time $t$ and depth $z$ (positive downwards). $w$ is the vertical velocity profile and $k$ is the thermal diffusivity profile of the the ice-sheet, since $k$ is temperature-dependent, the problem is non-linear. The calculation of diffusivity $k$ and velocity $w$ is detailed in Appendix \ref{fwd_appendix}. By convention, the forward model is defined over the temporal domain $t \in [t_0, t_{\mathrm{pr}}]$, representing heat transfer from $t_0$ years before present to the present ($t=t_{\mathrm{pr}}$), and over the spatial domain $z \in [0, H]$, where $z=0$ and $z=H$ denote the surface and the base of the ice sheet, respectively. We employ Dirichlet boundary conditions for the forward model simulations performed in this study, which are given by,
\begin{subequations} 
\label{seq_bcs} 
 \begin{flalign}
    &T_{\theta}(t,0)=\theta(t), \label{eq:subeq1}&\\
    &T_{\theta}(t,H)=\theta_{b}.
    \label{eq:subeq2}
\end{flalign}
\end{subequations}
The top boundary condition $T_{\theta}(t,0)$ is determined by the surface temperature history $\theta(t)$ and bottom boundary condition $T_{\theta}(t,H)$ is given by the constant basal temperature $\theta_{b}$ of the ice sheet.

The borehole has length $L$, typically extending a few hundred meters into the ice, with $L < H$. Given $\theta(t)$, the forward model simulates the temperature profile over the full vertical domain $[0, H]$, including the borehole and the ice beneath it. We discretize Equation (\ref{eq_hae}) using the forward Euler method in time, central differences for the diffusion term, and forward differences for the advection term. We use temporal resolution of $2^{-4} ~\mathrm{yr}$ and spatial resolution of $\sim4 ~\mathrm{m}$ for all forward simulations performed in this study. In order to determine the initial condition of the ice-sheet, the model is run with a given mean temperature $\theta_{m}$ and the basal temperature $\theta_{b}$ as the top and bottom boundaries, respectively, until the profile reaches equilibrium \cite{shaju2025sensor}. 

\subsection{Surface temperature models}
\label{sec:transdimensional}
The surface temperature model incorporates both recent climate changes and a pre-observational mean temperature $\theta_{pom}$ \cite{Beltrami2005longterm, verdoya2007pom} against which these changes can be referenced. We parametrize the surface temperature using a set of parameters $\vec{\Theta}$, which characterize the evolution of $\theta(t)$ through a surface temperature model $\Gamma$,
\begin{equation}
\label{surface_temp_model}
    \theta(t) = \Gamma(\vec{\Theta}). 
\end{equation}

$\theta(t)$ can be represented in many ways for the purpose of climate reconstruction. Studies have represented $\theta(t)$ as an adaptive piecewise-linear function \cite{Hopcroft2007inference, Hopcroft2023global,Muto2011recent} and \cite{Orsi2012little} used a linear combination of Fourier basis functions to set up $\theta(t)$. Finding the components (parameters) $\vec{\Theta}$ of the function that represent $\theta(t)$ is the main task performed by the inverse model. Therefore, $\theta(t)$ should be modeled in a way that balances resolution of $\theta(t)$ being reconstructed and complexity of inverse model, as increasing the number of components directly affects computational cost (Section \ref{subsec:mhmcmc}).

\subsubsection{Adaptive piecewise-linear model}
\label{sec:piecewise_linear}
We implement the adaptive piece-wise-linear model by \cite{Hopcroft2007inference, Muto2011recent} as the reference model. As per this model, the surface temperature history is represented as a piecewise-linear function, with nodes defined at 
$N_{AP}$ time points of surface temperature. The value of $N_{AP} \in \mathbb{N}$ is varied between a specified minimum $N_{AP_{min}}$ and maximum $N_{AP_{max}}$ number of nodes \cite{Hopcroft2007inference, muto2011multidecadal}. Let the time points be  $\{t_j\}_{j=0}^{N_{AP}} \subset [t_0, t_{pr}]$ such that  $0 = t_0 < t_1 < \cdots < t_{N_{AP}} = t_{pr}$, 
then, the piecewise-linear model can be written as,
\begin{equation}
\label{eqn:adaptive_piece_wise_linear}
\begin{aligned}
&\theta(t) =
\begin{cases}
\theta(t_j) + \dfrac{\theta(t_{j+1}) - \theta(t_j)}{t_{j+1} - t_j} (t - t_j), & \forall t \in [t_j, t_{j+1}],\ \forall j = 0, \dots, N_{AP} - 1.
\end{cases}
\end{aligned}
\end{equation}

\subsubsection{Kernel-based model}
\label{subsec:kernel_model} 
As an alternative, we
introduce a kernel-based surface temperature model, in which surface temperature variations are represented as a linear combination of kernel basis functions. A kernel or a covariance function, specifies the similarities between two values of a function evaluated at two input points \cite{duvenaud2014automatic}. In this context, the kernel specifies the covariance between temperatures at two time points.

A linear combination of weighted kernels allows us to represent surface temperature fluctuations by adjusting its weights. By adding the pre-observational mean temperature $\theta_{pom}$ to this fluctuations, we can construct the surface temperature time series,

\begin{equation}
\label{eqn:kernel_based_model}
\theta(t) = \theta_{pom} + \sum_{i=1}^{N_{KB}} \alpha_i \, \phi(t, t_i),
\end{equation}
where, $ \phi(t, t_i)$ is the kernel, $N_{KB} \in \mathbb{N}$ is the number of kernels, $\alpha_i$ is the weight of the $i^{th}$ kernel.
We use squared exponential (Gaussian) kernels $\phi(t,t_i)$, which are characterized by their centers $t_i$ and their widths $\gamma$ (also referred to as kernel length-scale),
\begin{equation}
\label{eqn:kernel_def}
    \phi(t,t_i) = \exp \left( -\frac{\|t - t_i\|^2}{2\gamma^2} \right).
\end{equation}

\section{Bayesian Inference for solving the inverse problem}
\label{sec:bayes}
In a Bayesian formulation, the unknown model parameters $\vec{\Theta}$ are assumed to follow a probability density based on our beliefs, known as the prior probability density $p(\vec{\Theta})$. $p(\vec{y}|\vec{\Theta})$ is the probability density (likelihood) for observing the data $\vec{y}$ given $\vec{\Theta}$. Bayes’ theorem states that for a particular realization of the data $\vec{y}$, the posterior density $p(\vec{\Theta} \mid \vec{y})$ can be written as,
\begin{equation}
\label{eqn_bayes}
     p(\vec{\Theta} \mid \vec{y}) = \frac{p(\vec{y} \mid \vec{\Theta})p(\vec{\Theta})}{p(\vec{y})}.
     \end{equation}
Bayesian inference solves the inverse problem by quantifying the posterior probability density of the unknown parameters, by addressing uncertainties in the observations, the forward model, and prior knowledge of the model parameters \cite{Bardsley2012mcmc,Martin2012A,lie2018random}. Note that we here neglect the uncertainty due to the discretization and choice of parameters in the forward heat transfer model.

\subsection{Likelihood}
\label{subsec:likelihood}

The likelihood represents the probability density for observing the data $\vec{y}$ for a given $\vec{\Theta}$, and is determined via Equation (\ref{eqn_dm_noise})  \cite{Bardsley2012mcmc}. Uncertainty in the observations is attributed to the measurement error $\vec{\epsilon}$, which is modeled in the Bayesian setting as an $n$-dimensional i.i.d.\ Gaussian random vector with variance $\sigma_m^2$. Hence, we quantify the likelihood as a multivariate Gaussian,
\begin{equation}
p(\vec{y} \mid \vec{\Theta}) = \frac{1}{\sqrt{(2\pi)^{n}\,\det{\mathbf{C_m}}}} \exp\!\left[-\frac{1}{2}(\vec{\hat{y}} - \vec{y})^{\top}
\mathbf{C_m}^{-1}(\vec{\hat{y}} - \vec{y})\right],
\end{equation}
where $\vec{y}$ denotes the observed borehole temperature measurements, and $\vec{\hat{y}}$ denotes the modeled borehole temperatures obtained by forward simulating the surface temperature history generated using the given $\vec{\Theta}$ (through $\Gamma(\vec{\Theta})$, Equation (\ref{surface_temp_model})) evaluated at the measurement points. $\mathbf{C_m}$ is an $n \times n$ diagonal matrix representing measurement error as uncorrelated Gaussian noise,
\begin{equation}
\label{measurement_error_matrix}
\mathbf{C_m} = \sigma_m^2\ \mathbf{I}.
\end{equation}

\subsection{Prior}
\label{subsec: prior}

We assume a vague prior for the surface temperature $\theta(t)$, with a normal distribution having  mean $\theta_m$ and standard deviation $\sigma_\theta$, at each point in time. Parameters $\vec{\Theta}$ of the surface temperature models (Section \ref{sec:piecewise_linear}-\ref{subsec:kernel_model}) which represent $\theta(t)$, have their own associated prior distributions. 

In the adaptive piecewise-linear model, the time points $\{t_j\}_{j=0}^{N_{AP}}$ are drawn using
order statistics from a uniform distribution over the time interval between $t_0$ and $t_{\mathrm{pr}}$ \cite{green1995reversible, Hopcroft2007inference, muto2011multidecadal}, and the prior on the corresponding temperature values $\{\theta(t_j)\}_{j=0}^{N_{AP}}$ is a uncorrelated multivariate Gaussian centered around the known present day mean temperature $\theta_m$ with a variance of $\sigma_{\theta}^2$ \cite{ Hopcroft2007inference}. Further details on prior are provided in Appendix \ref{rjmcmc_appendix}. 

In the kernel-based model, we select $\alpha_i \sim \mathcal{N}(0,\sigma_{\alpha}^2\mathbf{I}), i =1, \dots,N_{KB}$, where $\sigma_{\alpha}^2$ is chosen such that, for the specified values of $N_{KB}$ and $\gamma$, the model generates surface temperature time series with variance $\sigma_\theta^2$, at every point in time. To achieve a pointwise standard deviation $\sigma_{\theta}=1 ~\mathrm{K}$, $\sigma_{\alpha}$ is set as $0.6$ for a kernel configuration of $N=40$ and $\gamma=20$, and $\sigma_{\alpha}$ is set as $0.49$ for a kernel configuration of $N=60$ and $\gamma=20$. It is observed that, for the same time interval $(D)$, increasing the number of equidistantly spaced kernels, requires a comparatively smaller $\sigma_{\alpha}$ to attain the same specified value of $\sigma_{\theta}$. To achieve $\sigma_{\theta}=2~\mathrm{K}$, $\sigma_{\alpha}$ is set as $1.2$ for a kernel configuration of  $N=40$ and $\gamma=20$. We choose $\theta_{pom} \sim \mathcal{U}(\theta_{min}, \theta_{max})$.

\subsection{Markov Chain Monte Carlo sampling}
\label{subsec:mhmcmc}
For complex, non-linear forward models and high-dimensional parameter spaces, the posterior density $p(\vec{\Theta} \mid \vec{y}) $ cannot be computed analytically. This is because it requires evaluating 
$p(\vec{y})$, the marginal likelihood of the data (or evidence), which is computationally challenging in high-dimensional models as it involves computation of integrals over the entire parameter space \cite{Cai2022Proximal, Hopcroft2007inference, Groenke2023thermalstate}. Therefore, we approximate the posterior distribution using Markov chain Monte Carlo (MCMC) sampling.

It is a procedure of sampling from the desired distribution through generating a random walk in the parameter space \cite{Foreman_Mackey_2013emcee}. In a Markov chain, each point $S(i)$ represents the $i^{th}$ sample collected for $\vec{\Theta}$ and is solely determined by the value of the sample collected in the preceding step $S(i-1)$ \cite{Foreman_Mackey_2013emcee}. The Metropolis–Hastings algorithm is the simplest MCMC algorithm (MH-MCMC). The procedure for a single MH-MCMC iteration is formulated in Algorithm \ref{alg_mhmcmc}. From a given position $S(i)$, the MH-MCMC algorithm proceeds by drawing sample $S^*$ from the proposal density $Q(S^*;S(i))$ and accepts it with probability,  
\begin{equation}
\label{eqn_mh_ratio}
q = min \left( 1 , \frac{p(S^*)}{p(S(i))}\frac{p(\vec{y}\mid S^*)}{p(\vec{y}\mid S(i))}\frac{Q(S(i);S^*)}{Q(S^*;S(i))}\right).
\end{equation}
$Q(S^*;S(i))$ is the proposal density centered around $S(i)$, $p(S^*)$ is the prior, and $p(\vec{y}\mid S^*)$ is the likelihood of the proposed sample $S^*$. Similarly, $Q(S(i);S^*)$ is proposal density centered around $S^*$, while $p(S(i))$ and $p(\vec{y}\mid S(i))$ are the prior and likelihood, respectively, of the sample at the current position $i$. 

\begin{algorithm}
\caption{A single MH-MCMC iteration}
\label{alg_mhmcmc}
\begin{algorithmic}[1]
\STATE{Draw $S^* \sim  Q(S^*;S(i))$}
\STATE{Calculate $q$ as per Equation (\ref{eqn_mh_ratio})}
\STATE{Draw $u \sim \mathcal{U}(0,1)$}
\IF{$u \le q$}
\STATE{$S(i+1) = S^*$}
\ELSE
\STATE{$S(i+1) = S(i)$}
\ENDIF
\end{algorithmic}
\end{algorithm}

 The MH-MCMC converges (as $i \rightarrow \infty$) to a set of samples which represent the posterior distribution. It can sample a fixed-length parameter vector $\vec{\Theta}$, but it takes longer to converge if there are many parameters (in $\vec{\Theta}$). 

 The central idea of convergence of the MCMC sampler is to collect a finite set of approximately independent samples for approximating a stationary distribution representing posterior (i.e., the target distribution). A sampler that satisfies detailed balance leaves the target distribution invariant; together with ergodicity, this ensures that the chain converges to the posterior, from which a finite set of approximately independent samples can then be collected \cite{Barrosdetailedbalance2023, reich2015probabilistic}.  

\subsubsection{Reversible Jump Markov Chain Monte Carlo sampler}
\label{subsec:rjmcmc} 
Reversible jump Markov Chain Monte Carlo (RJ-MCMC) is an enhanced form of MH-MCMC to have a varying-length parameter vector $\vec{\Theta}$. It has the capability to jump between parameter subspaces of differing dimensionality \cite{green1995reversible}. This procedure is useful when the number of parameters is itself unknown as in the adaptive piecewise-linear surface temperature model. 

 At each iteration, a sample of surface temperature model parameters is drawn from a proposal distribution associated with updating a parameter value, such as updating a time point ($t_j$ in Equation (\ref{eqn:adaptive_piece_wise_linear})), updating a temperature value, updating the pre-observational mean, or adding or deleting a time point along with its corresponding temperature value. The dimensions of the parameter space thus varies in relation to the number of the surface temperature time-points. The procedure for a single RJ-MCMC iteration is described in Algorithm \ref{alg_rjmcmc}. At each iteration, the sampler randomly selects a type of move and updates the current sample $S(i)$ of the surface temperature history accordingly. The resulting proposed sample $S^*$,  incorporates the selected move, with the associated change drawn from the corresponding proposal distribution. The details on prior, proposals, Jacobian calculations, and acceptance criteria as per \cite{Hopcroft2007inference, muto2011multidecadal} are detailed in Appendix \ref{rjmcmc_appendix}. 
The Jacobian accounts for the transformations between parameter spaces of different dimension; for moves that leave the dimension unchanged, it equals $1$ \cite{Hopcroft2007inference}.  

\begin{algorithm}
\caption{A single RJ-MCMC iteration.}
\label{alg_rjmcmc}
\begin{algorithmic}[1]

\IF{$N_{AP_{\min}}<N_{AP}<N_{AP_{\max}}$} 
    \STATE{$ch \sim \mathrm{Unif}\{1,2,3,4,5\}$}
\ELSIF {$N_{AP} =N_{AP_{\min}}$}
    \STATE{$ch \sim \mathrm{Unif}\{1,3\}$}
\ELSIF {$N_{AP} = N_{AP_{\max}}$}
    \STATE{$ch \sim \mathrm{Unif}\{1,2,4\}$}  
\ENDIF  

\COMMENT{Select the proposal density based on the chosen move $ch$.}
\IF{$ch = 1$}
    \STATE {\text{$Q(S^*;S(i))$ $\gets$ Proposal density for modifying $\theta(t_j)$ at random position $j$.} }
\ELSIF{$ch = 2$}
    \STATE {\text{$Q(S^*;S(i))$ $\gets$ Proposal density for modifying $t_j$ at random position $j$.} }
\ELSIF{$ch = 3$}
    \STATE {\text{$Q(S^*;S(i))$ $\gets$ Proposal density for adding a new point $(t_j,\theta(t_j))$}. \\
    $N_{AP}\leftarrow N_{AP}+1$.}

\ELSIF{$ch = 4$}
    \STATE {\text{$Q(S^*;S(i))$ $\gets$ Proposal density for deleting an existing point $(t_j,\theta(t_j))$}. \\
    $N_{AP}\leftarrow N_{AP}-1$.} 

\ELSIF{$ch = 5$}
    \STATE {\text{$Q(S^*;S(i))$ $\gets$ Proposal density for modifying $\theta_{pom}$.} }
\ENDIF

\STATE{$S^* \sim  Q(S^*;S(i))$}

\STATE {
$q \leftarrow min \left( 1 , \frac{p(S^*)}{p(S(i))}\frac{p(\vec{y}\mid S^*)}{p(\vec{y}\mid S(i))}\frac{Q(S(i);S^*)}{Q(S^*;S(i))}\det J\right)$
}

\STATE{$u \sim \mathcal{U}(0,1)$}
\IF{$u \le q$}
    \STATE{$S(i+1) = S^*$}
\ELSE
    \STATE{$S(i+1) = S(i)$}
\ENDIF
\end{algorithmic}
\end{algorithm}

\textit{RJ-MCMC sampler implementation:}
The RJ-MCMC sampler is implemented as per \cite{Hopcroft2007inference} and the overall structure of the implementation is outlined in Algorithm \ref{alg_rjmcmc}. The  hyperparameters involved were chosen as per \cite{muto2011multidecadal} and are detailed in the Appendix \ref{rjmcmc_appendix}.

\subsubsection{Affine-invariant ensemble MCMC sampler}
\label{subsec:emcee} 
The affine-invariant ensemble sampling algorithm proposed by \cite{GoodmanWeare2010ensemble} is based on the principle of the stretch move and employs multiple interacting Markov chains that run simultaneously to sample the target distribution. Each chain is generated by a walker, where a walker is one member of the ensemble of samplers.

We start by summarizing the serial stretch move algorithm as described in  \cite{Foreman_Mackey_2013emcee, GoodmanWeare2010ensemble}. We assume an ensemble of $K$ walkers as $\mathbb S = \{S_k\}_{k=1}^K$ evolving simultaneously. The proposal distribution for one walker $k$ is determined by the current positions of the remaining $K-1$ walkers. These remaining walkers constitute the complementary ensemble $\mathbb S_{[k]}=\{S_j \mid \forall j \neq k\}$. To update the position (i.e.,  positions for all $N_{\Theta}$ parameters) of a walker $S_k(i)$, we sample  $S_k^*$ as:
\begin{equation}
\label{emcee_proposal}
S_k^* = S_j + R[S_k(i) - S_j] 
\end{equation}
where $R$ is a random variable drawn according to a density $g(r)$. If this density $g$ of the scaling variable $R$ satisfies $g(1/r)=rg(r)$, then proposal distribution in Equation (\ref{emcee_proposal}) is symmetric \cite{GoodmanWeare2010ensemble, Foreman_Mackey_2013emcee}. It is recommended to use the following form of $g(r)$,
\begin{equation}
\label{scaling_element}
g(r) \propto
\begin{cases}
\dfrac{1}{\sqrt{r}}, & \text{if } r \in \left[\dfrac{1}{a},\, a\right], \\
0, & \text{otherwise}.
\end{cases}
\end{equation}
where the parameter $a > 1$ is a tunable parameter that can be adjusted to improve performance \cite{GoodmanWeare2010ensemble}. As shown by \cite{GoodmanWeare2010ensemble}, the conditional density associated with the straight-line stretch move (for all $N_{\Theta}$ parameters) is proportional to,
\begin{equation}
\label{conditional}
\lVert S_k^* - S_j\rVert ^{N_{\Theta}-1} \pi(S_k^*),  
\end{equation}
where $\pi$ represents the probability density.

Since the proposal is symmetric (Equation (\ref{emcee_proposal})), the chain satisfies detailed balance if the proposal is accepted with probability,
\begin{equation}
\label{emcee_acceptance}
q  = min\left( 1, R^{N_\Theta-1} \frac{\pi(S_k*)}{\pi(S_k(i))}\right),
\end{equation}
where $N_\Theta$ is the dimension of the parameter space \cite{GoodmanWeare2010ensemble, Foreman_Mackey_2013emcee}. This procedure is then repeated for each walker in the ensemble in series.

\cite{Foreman_Mackey_2013emcee} proposes a parallel version of the serial stretch move in which walkers are advanced
simultaneously. This is done by splitting the ensemble of all walkers into two subsets: $\mathbb S_{[0]} =\{S_k \mid k = 1, ..., \frac{K}{2}\}$ and $\mathbb S_{[1]}=\{S_k,\mid k = \frac{K}{2} +1, ..., K\}$. All walkers in $\mathbb S_{[0]}$ are simultaneously updated using the stretch move procedure described in Equation (\ref{emcee_proposal}) such that $S_j \in \mathbb S_{[1]}$. Then, using the new positions $\mathbb S_{[0]}$, positions of $\mathbb S_{[1]}$ are updated and thus providing a valid step for all walkers. This procedure by \cite{Foreman_Mackey_2013emcee} is described in Algorithm \ref{alg_emcee}.

\begin{algorithm}
\caption{A single step of parallel stretch move}
\label{alg_emcee}
\begin{algorithmic}[1]
\FOR{ $s \in \{0,1\}$} 
\FOR{$k \in \{1, ..., \frac{K}{2}\}$} 
\STATE{Draw a walker $S_j$ from the complementary ensemble $\mathbb S_{[\tilde  s]}(i)$.}
\STATE{$S_k \leftarrow \mathbb S_{[s]}[k]$}
\STATE{$r \leftarrow R \sim g(r)$}
\STATE{$ S_k^* \leftarrow S_j + R[S_k(i) - S_j]$}

\STATE{$q  \leftarrow R^{N_\Theta-1} \frac{p(S_k^*)}{p(S_k(i))}$}
\STATE{$u \sim \mathcal{U}(0,1)$}
\IF{$u \le q$}
\STATE{$S_k(i+\frac{1}{2}) = S_k^*$}
\ELSE
\STATE{$S_k(i+\frac{1}{2}) = S_k(i)$}
\ENDIF
\ENDFOR \COMMENT{This loop can be run in parallel for all $k$.}
\STATE{$i \leftarrow i+\frac{1}{2}$}
\ENDFOR
\end{algorithmic}
\end{algorithm}

We used \texttt{emcee}, a Python implementation of the affine-invariant ensemble Markov chain Monte Carlo (MCMC) sampler of \cite{GoodmanWeare2010ensemble}, developed by \cite{Foreman_Mackey_2013emcee} to sample kernel-based surface temperature model parameters.

\subsection{Posterior approximation}
\label{subsec:posterior_approx_conv}
The convergence of the RJ-MCMC sampler used to approximate the posterior distribution of the piecewise-linear model is assessed by continuing the simulation until the posterior estimates (Section \ref{subsec:estimates}) stabilize and remain unchanged upon further addition of the sampled surface temperature time series  \cite{Hopcroft2007inference, Muto2011recent}.
For the \texttt{emcee} sampler used to approximate the posterior distribution of the kernel-based surface temperature model, the convergence is diagnosed using the autocorrelation time. The autocorrelation time measures the number of evaluations of the posterior density function required to generate independent samples \cite{Foreman_Mackey_2013emcee}. The \texttt{emcee} sampler computes the autocorrelation time for all parameters. Following the heuristic guideline provided in \cite{Foreman_Mackey_2013emcee}, we use a total sample size of approximately 50 times the maximum autocorrelation time across all parameters and discard a burn-in period of approximately 10 times the same.
The number of samples used by the samplers for the experiments are listed in their respective sections.

\subsection{Posterior statistics}
\label{subsec:estimates} 
We summarize the approximated posterior by a point estimate and a pointwise credible interval, defined as follows for both surface
temperature models.

\subsubsection{Point estimates}
\label{point_estimates}
We use the posterior mean as the point estimate. Because the kernel-based model is linear in its weights, the time series generated from the posterior
mean of the parameters coincides with the mean of the sampled surface temperature histories. For the adaptive piecewise-linear model, whose parameter vector has varying
length, the posterior mean is obtained by averaging the sampled histories at each time
point \cite{Hopcroft2007inference}.

\subsubsection{Pointwise credible intervals}
\label{uncertainty_bar}
In addition to point estimates, we also require ranges that indicate where the surface temperature histories are expected to lie. We use the concept of credible interval in Bayesian statistics which is defined as $100(1-a)$ for a given univariate posterior density \cite{murphy2012machine}. Here,  the posterior uncertainty is quantified using pointwise credible intervals obtained by removing $\frac{a}{2}\%$ from both upper and lower ends representing highest and lowest of total surface temperature time series samples  at each time point. We show the $95 \%$ pointwise credible interval for both surface temperature models.

\section{Surface temperature reconstruction using kernel-based model}
\label{sec:reconstruction using kbm}
\begin{figure}[h]
  \centering
  \label{fig:fig_art_data}
  \includegraphics[width=\textwidth]{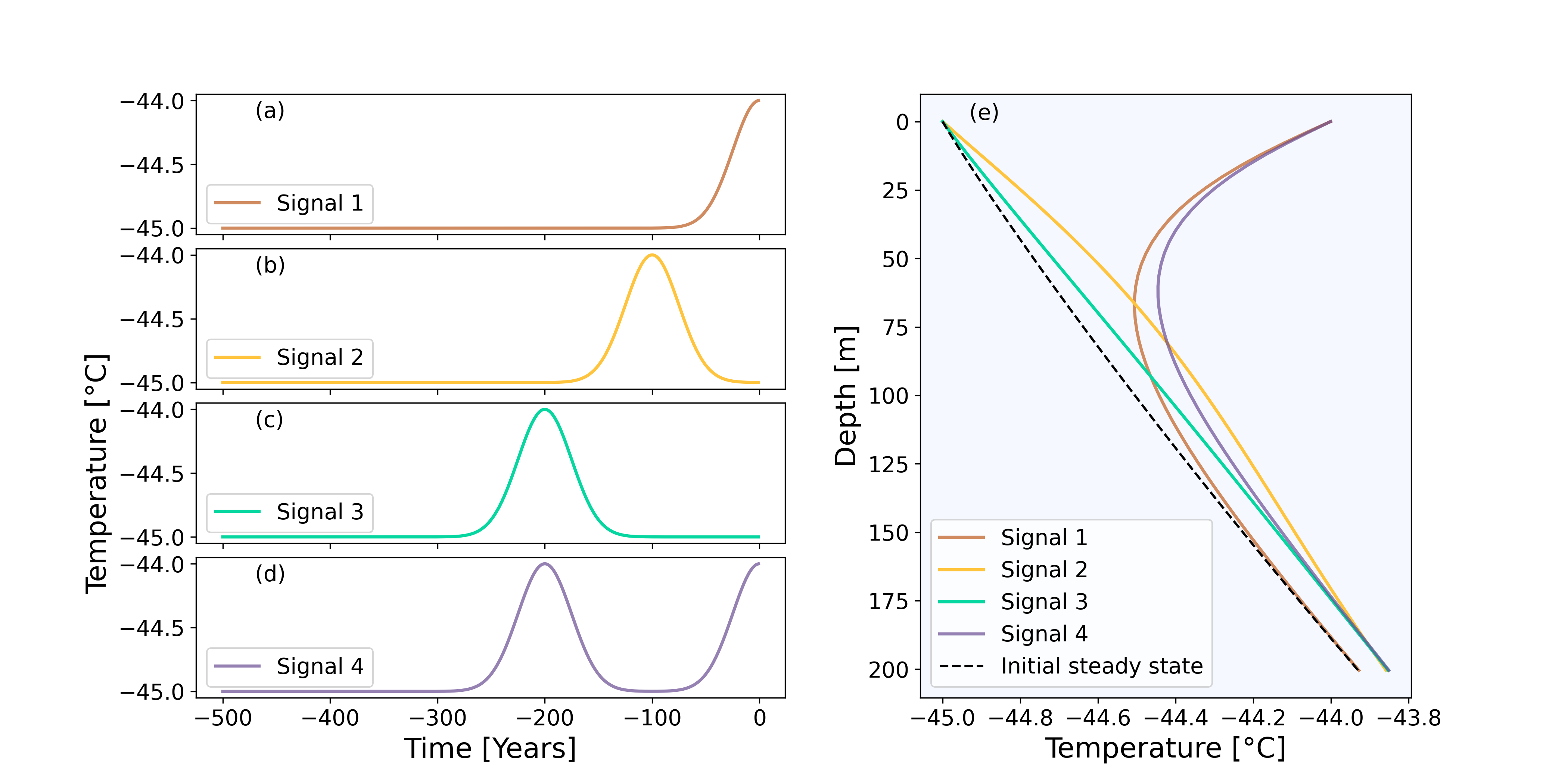}
  \caption{Synthetic surface temperature signals and the borehole temperature profiles. (a)-(d) show synthetic surface temperature signals centered at various points in time. (e) shows their corresponding forward-model simulations (top 200 m), from which borehole measurements are extracted. The black dashed line in (e) represents the initial condition used in the forward simulations.}
\end{figure}

We evaluate the kernel-based surface temperature model on four synthetic surface temperature histories (Fig.~\ref{fig:fig_art_data}a–d). Borehole observations are created by forward-simulating each history (Fig.~\ref{fig:fig_art_data}e) and
extracting temperatures from the resulting profile at the measurement depths.

Signal 1 shows a warming trend toward the present, Signals 2 and 3 represent the occurrence of events at 100 and 200 years ago, respectively, and Signal 4 represents the occurrence of an event in the past together with a warming trend toward the present. We use the parallel ensemble sampler presented in Section \ref{subsec:emcee} to estimate kernel weights and pre-observational mean temperature. The priors are selected as $\theta_{\mathrm{pom}} \sim \mathcal{U}(-50, -40)$, and $\alpha_i \sim \mathcal{N}(0, \sigma_\alpha^{2}\mathbf{I})$-- $\sigma_\alpha$ is chosen such that standard deviation of surface temperature $\sigma_\theta $ is $1 ~\mathrm{K}$ (Section \ref{subsec: prior}).

For a given set of surface temperature time series, we start by discussing how to identify the kernel setup; that is the required number of kernels and its length-scale for efficient reconstructions. The surface temperature reconstructions are performed using the identified kernel setup. We investigate the influence of number of kernels, measurement uncertainty, measurement density, and temporal smearing on reconstruction. Finally, we compare the proposed approach with the conventional adaptive piecewise linear model employing RJ-MCMC sampler.

\subsection{Identifying the kernel setup}
\label{sec: kernel_set_up_identification}
A larger number of kernels with shorter length-scale is ideal for capturing small variations of a given time series. When it comes to the inverse problem, the larger the number of kernels, the higher is the computational cost due to the growing dimensionality of the parameter space.

\begin{figure}[htbp]
  \centering
  \label{fig:fig_kernel_setup}
  \includegraphics[width=\textwidth]{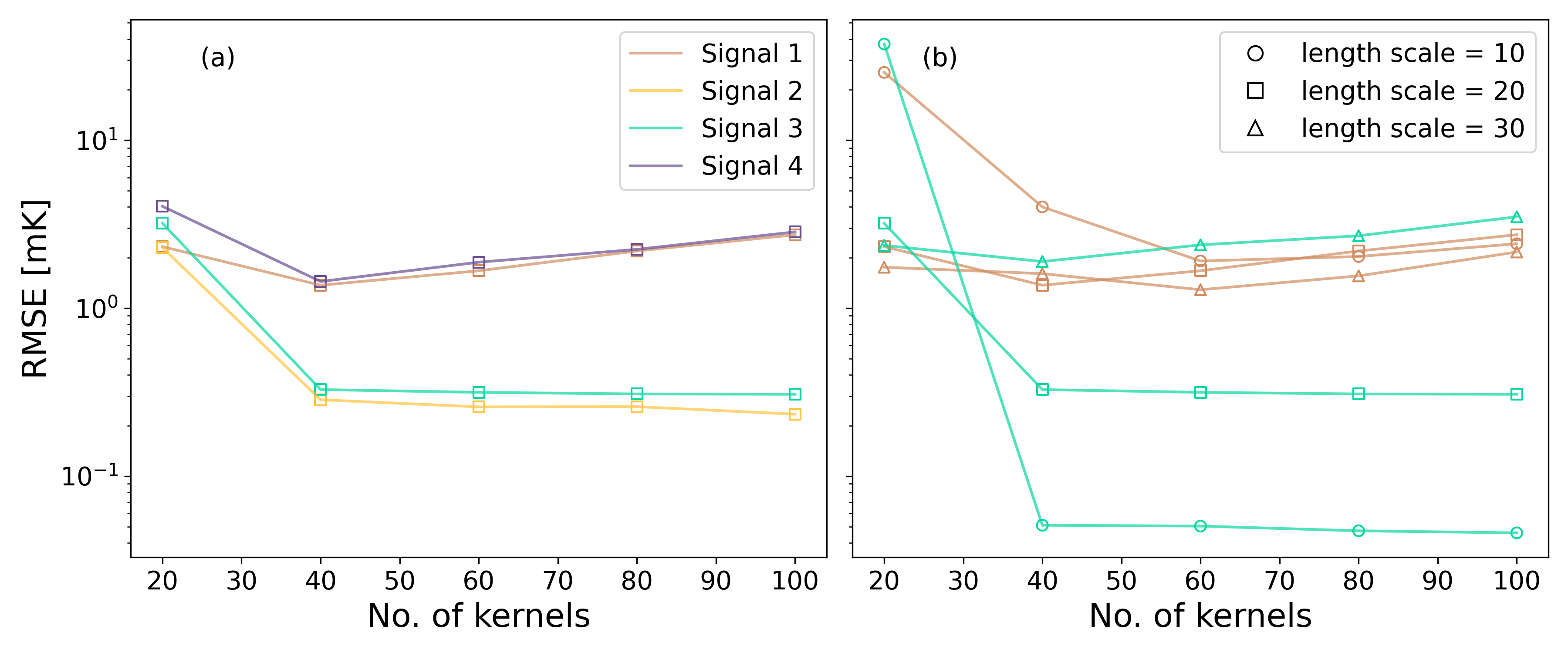}
  \caption{Error (as RMSE) in synthetic signal representations for different kernel configurations. (a) shows how the error varies with increasing number of kernels of length-scale 20, for all signals. (b) shows how the error varies with an increasing number of kernels for different length-scales, for Signal 1 and Signal 3.}
\end{figure}

To determine the appropriate number of kernels, and their length-scale, we examine how well various signals (Fig.~\ref{fig:fig_art_data}a-d) can be represented by a kernel setup. For this analysis, kernel weights are optimized directly against the synthetic signals using an optimization based on an $L2$-based loss function rather than inferred through Bayesian inversion.  The error quantified by the root mean squared error (RMSE), observed for representing signals using various kernel setup with its respective original signals show minimal variations beyond 40 kernels. For a fixed length-scale of 20, the warming signal exhibits a slight increase in error, whereas the remaining signals show a slight decrease as the number of kernels increases (Fig.~\ref{fig:fig_kernel_setup}a).  As the length-scale decreases, the error decreases slightly for signals without a warming trend beyond 40 kernels; however, this trend is not observed for signals with a warming trend (Fig.~\ref{fig:fig_kernel_setup}b). In this study, we choose a length-scale of 20. We see that, once a high enough number of kernels has been chosen, the error almost remains constant with very slight increment or decrement with increase in number of kernels (Fig.~\ref{fig:fig_kernel_setup}). Therefore, for Gaussian kernels (Equation (\ref{eqn:kernel_def})) with length-scale 20, we pick only 40 kernels to represent the aforementioned signals, and we use this kernel configuration throughout this study. To assure that this choice is still valid in the inverse problem setup, we compare the 40- to 60-kernel configuration in the inverse problem in the next section.

\subsection{Sensitivity of reconstructions to the number of kernels}
\label{sensitivity_analysis_kerne_no}
We reconstruct the signals (Signal 1--Signal 4) using kernel setups with 40 and 60 kernels, each with a length-scale of 20. For the 40-kernel configuration, we use $\sigma_\alpha = 0.6$, while for the 60-kernel configuration we use $\sigma_\alpha = 0.49$, as the prior setting (Section \ref{subsec: prior}). We collected samples using parallel affine invariant ensemble sampler (\texttt{emcee}), with the number of walkers set to $122$ for both configurations. For the configuration with 40 kernels, we generated $120,000$ samples (per walker), of which the first 20,000 samples (per walker) were discarded as burn-in. For the configuration with 60 kernels, we generated $240,000$ samples (per walker), with the first 40,000 samples (per walker) discarded as burn-in. 

\begin{figure}[htbp]
  \centering
  \label{fig:fig_kernel_sensitivity}
  \includegraphics[width=\textwidth]{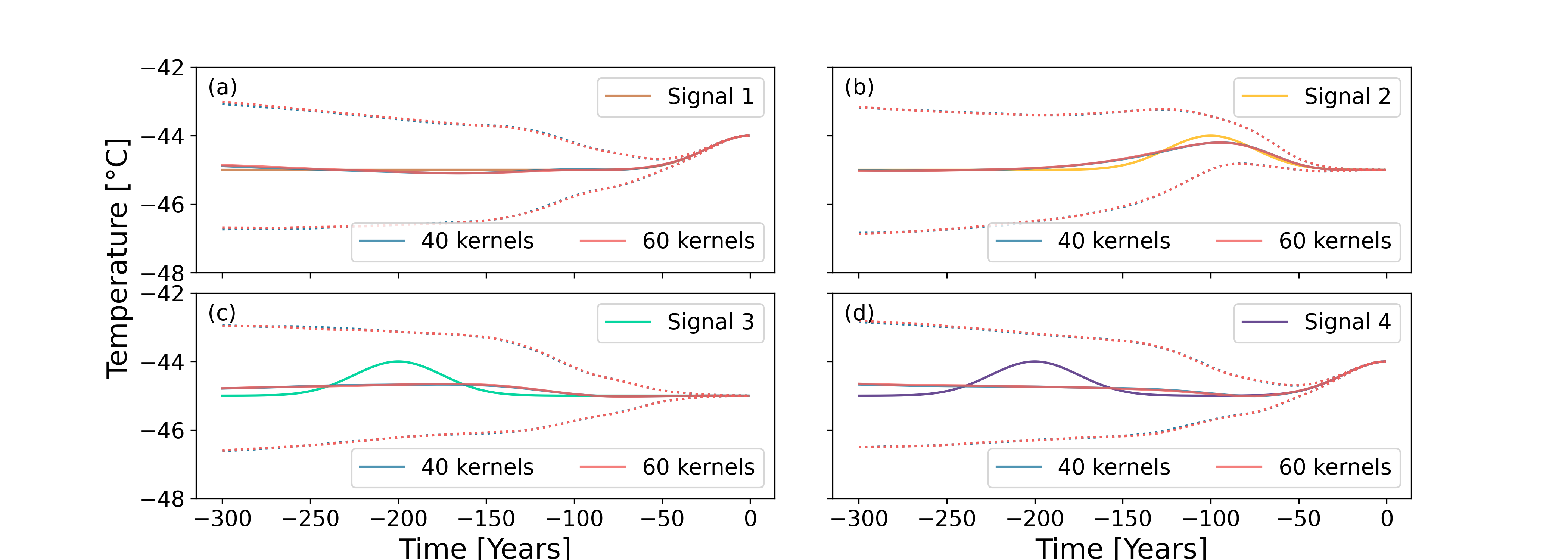}
  \caption{Synthetic signal reconstructions with varying number of kernels. Panels  (a),(b),(c) and (d) show signal reconstructions obtained when signals are represented with 40 (blue), and 60 (red) number of kernels. Solid lines denote the posterior mean, and dotted lines represent the corresponding pointwise credible interval.}
\end{figure}

In general, quality of reconstruction depends on how far back in time the signals originate. Signals 3 and 4 (Fig.~\ref{fig:fig_kernel_sensitivity}c,d) show higher errors for reconstructing their components from 200 years in the past, when compared to Signal 2 (Fig.~\ref{fig:fig_kernel_sensitivity}b) for its component from 100 years in the past. The uncertainty of reconstruction (represented by pointwise credible intervals) increases from the present day toward the past, as the amount of information from boreholes decreases (Fig.~\ref{fig:fig_kernel_sensitivity}a, b, c, and d).

The reconstructed signals obtained from both setups are nearly indistinguishable, indicating minimal difference. No systematic increase or decrease in reconstruction errors is observed for the 60-kernel configuration when compared to the 40-kernel configuration. Additionally, the associated point wise credible intervals significantly overlap. Thereby we conclude that reconstructions are not sensitive to increasing number of kernels, beyond 40 kernels.

\subsection{Reconstruction quality: measurement density vs measurement uncertainty}
\label{subsec:merr}
\begin{figure} [h]
  \centering
  \label{fig:fig_no_of_obs}
  \includegraphics[width=\textwidth]{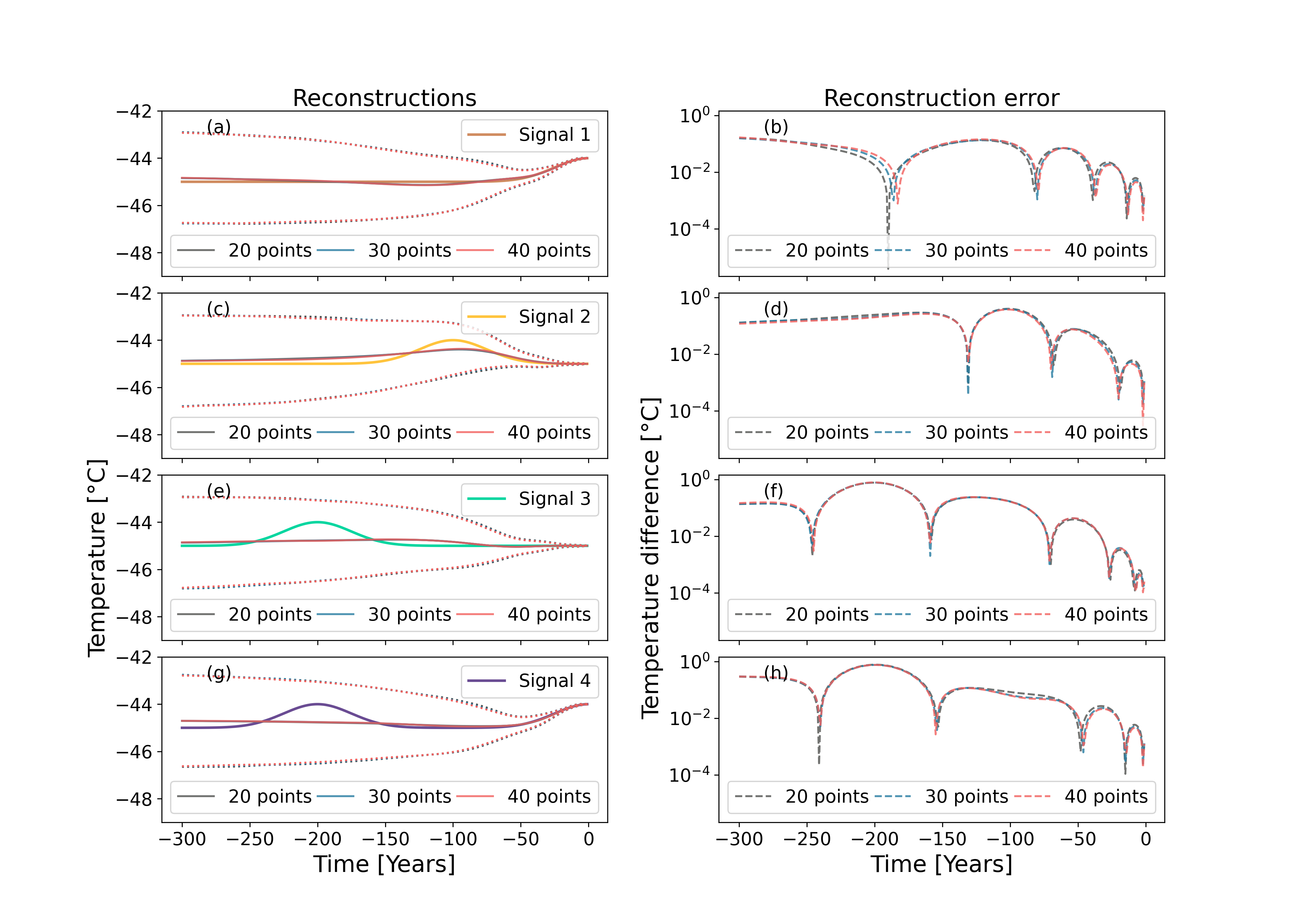}
  \caption{Synthetic signal reconstructions with different number of measurement points. Panels (a), (c), (e), and (g) show signal reconstructions obtained using 20 (gray), 30 (blue), and 40 (red) points. Solid lines denote the posterior mean, and dotted lines represent the corresponding pointwise credible interval. (b),(d),(f) and (h) shows the difference between posterior mean and the synthetic signal.}
\end{figure} 
To investigate the impact of measurement density on the synthetic signal reconstructions (Fig.~\ref{fig:fig_no_of_obs}a,c,e, and g), we use 20, 30, and 40 equidistantly placed points along the borehole from $0 ~\mathrm{m}$ to $200 ~\mathrm{m}$. Accordingly, the corresponding borehole temperature measurements are extracted from forward simulations (Fig.~\ref{fig:fig_art_data}e) of the signals at these sets of depth points. The standard deviation of  measurement uncertainty $\sigma_m$ is set to $10 ~\mathrm{mK}$ for analyzing the impact of measurement density. The reconstructed signals with all the three sets of measurement point settings aforementioned, overlaps each other for all signals (Fig.~\ref{fig:fig_no_of_obs}a,c,e and g) and so does their reconstruction error (Fig.~\ref{fig:fig_no_of_obs}b,d,f, and h). Also, reconstructions with high measurement density show only a very minor drop in their credible intervals (Fig.~\ref{fig:fig_no_of_obs}a,c,e, and g). Therefore, we infer that using denser measurements, which effectively reduces the spacing between them, does not significantly improve the reconstruction quality for the selected measurement uncertainty of $10~\mathrm{mK}$.

\begin{figure}[htbp]
  \centering
  \label{fig:fig_merr}
  \includegraphics[width=\textwidth]{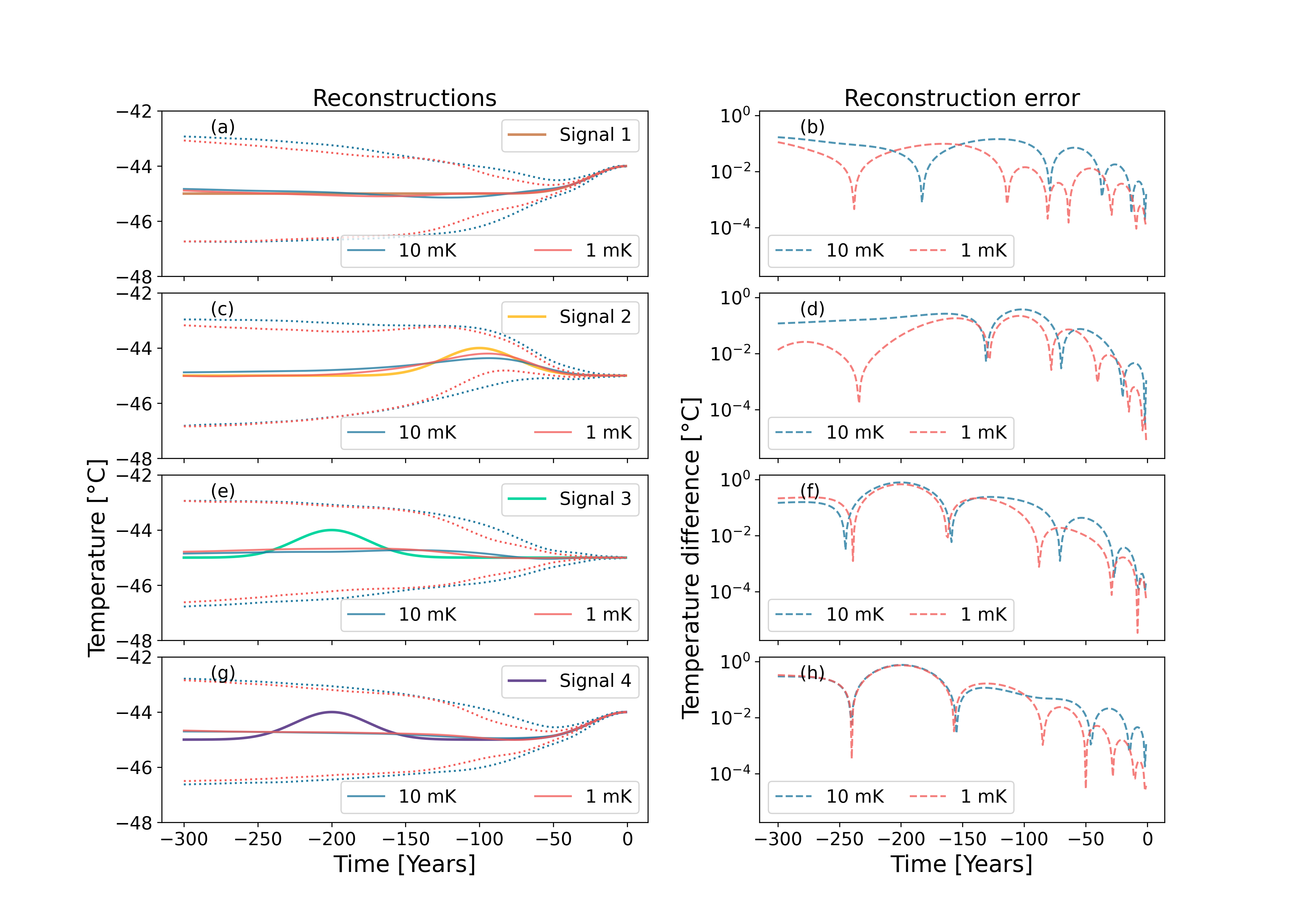}
  \caption{Synthetic signal reconstructions with varying measurement uncertainty. Panels (a), (c), (e), and (g) show signal reconstructions obtained using measurement uncertainty with $\sigma_m$ of $10 ~\mathrm{mK}$ (blue), and $1 ~\mathrm{mK}$ (red). Solid lines denote the posterior mean, and dotted lines represent the corresponding pointwise credible interval. (b),(d),(f) and (h) shows the difference between posterior mean and the synthetic signal.}
\end{figure}

We further investigate the performance of synthetic signal reconstructions with measurement uncertainty of standard deviations $\sigma_m =$  $10 ~\mathrm{mK}$, and $1 ~\mathrm{mK}$ (Fig.~\ref{fig:fig_merr}). The reconstructions are performed with a constant measurement points setting-- 40 equidistantly spaced measurement points between 0 and 200~m for all signals depicted in Fig.~\ref{fig:fig_merr}. A decreasing trend in reconstruction error is observed for all signals over the past century as the measurement uncertainty decreases (Fig.~\ref{fig:fig_merr}b,d,f,h). Reconstructions with more accurate measurements thus show a significant reduction in the reconstruction uncertainty (Fig.~\ref{fig:fig_merr}a,c,e,g). 

Therefore, reconstruction quality can be better improved by reducing the borehole temperature measurement uncertainty than by reducing the spacing between the measurement points (Fig.~\ref{fig:fig_no_of_obs} and Fig.~\ref{fig:fig_merr}), and this is in accordance with the conclusions of \cite{Clow_1992}.

\subsection{Influence of borehole depth and temporal smearing}

\begin{figure}[htbp]
  \centering
  \label{fig:temporal_smearing}
  \includegraphics[width=\textwidth]{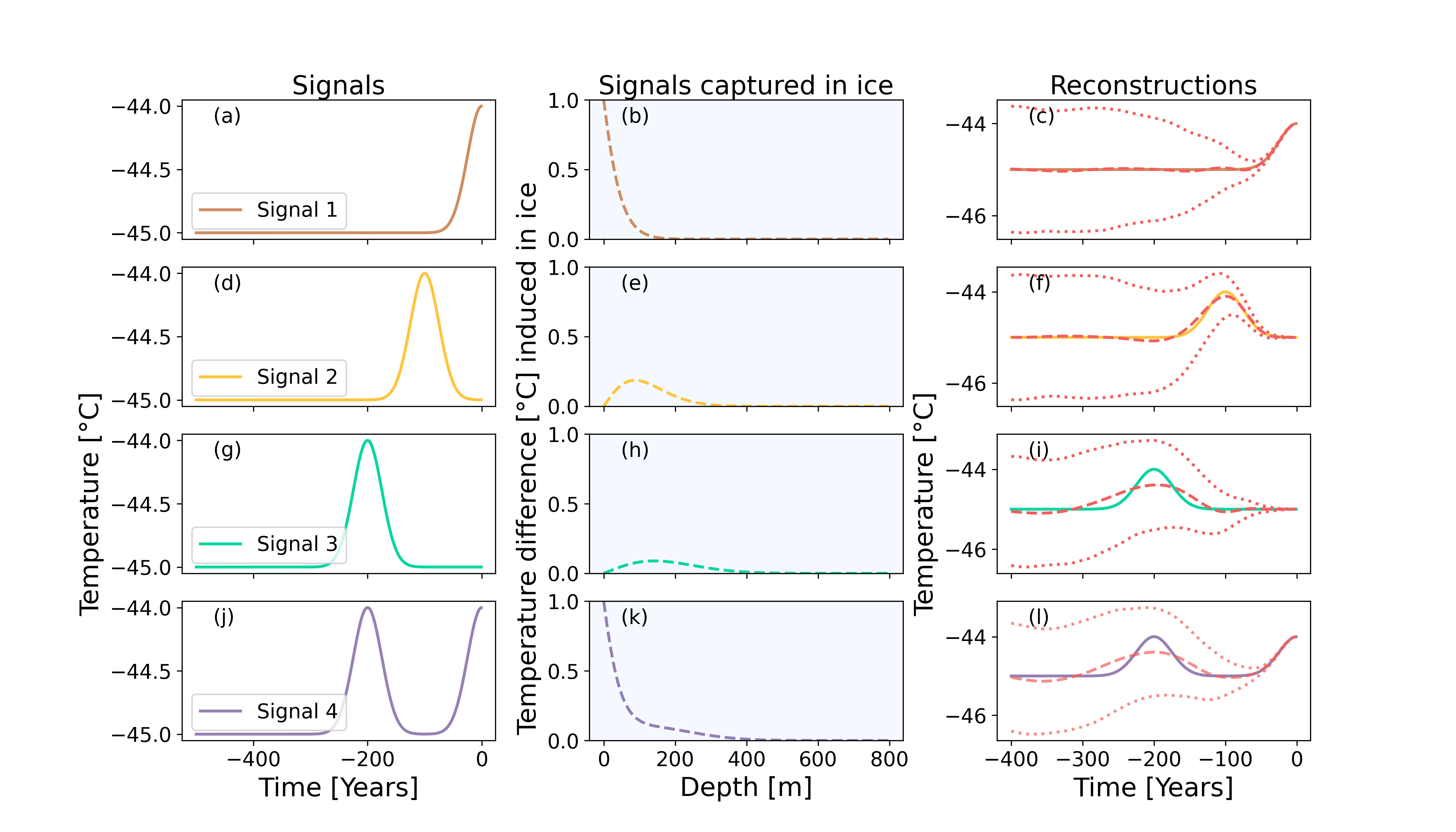}
    \caption{Synthetic signal reconstructions using the temperature profile of the entire ice column. (a), (d), (g), and (j) show the signals. (b), (e), (h), and (k) show the strength of the signals captured in the ice, obtained by subtracting the initial condition from the respective forward simulations of the signals. (c), (f), (i), and (l) show the reconstructions using  measurements as the forward simulated temperature profile of the the entire ice column.}
\end{figure}

To analyze the influence of borehole depth and temporal smearing (i.e., the attenuation of signals recorded in ice over time) on the reconstruction, we repeat the inversion using temperature measurements from the full ice column ($2782~\mathrm{m}$ thick), assuming a measurement uncertainty of $\sigma_m = 1~\mathrm{mK}$ (Fig.~\ref{fig:temporal_smearing}).

The quality of the reconstructions using full ice column measurements is exceptionally good compared to the shallow ice borehole based reconstructions with identical measurement uncertainty (Fig.~\ref{fig:fig_merr} and Fig.~\ref{fig:temporal_smearing}). This is because the measurements from the full ice column fully capture the information about the induced changes in heat distribution caused by Signals 1–4 (Fig.~\ref{fig:temporal_smearing}b,e,h,k). Furthermore, Signals 1 and 2, corresponding to events from the recent present and 100 years ago, are reconstructed nearly perfectly (Fig.~\ref{fig:temporal_smearing}c,f), while the signals containing the 200-year event remain slightly distorted (Fig.~\ref{fig:temporal_smearing}i,l) in the full ice column based reconstructions, demonstrating the influence of temporal smearing. Comparing both effects, the reconstructions are more strongly affected by missing depth coverage than by signal attenuation alone.

\subsection{Comparison with the adaptive piecewise-linear approach}

To place the proposed kernel-based framework in the context of existing Bayesian inversion approaches, we compare reconstructions obtained using the reference adaptive piecewise-linear model sampled with RJ-MCMC with those obtained using the kernel-based model.

\begin{figure}[htbp]
  \centering
  \label{fig:fig_rjmcmc}
  \includegraphics[width=\textwidth]{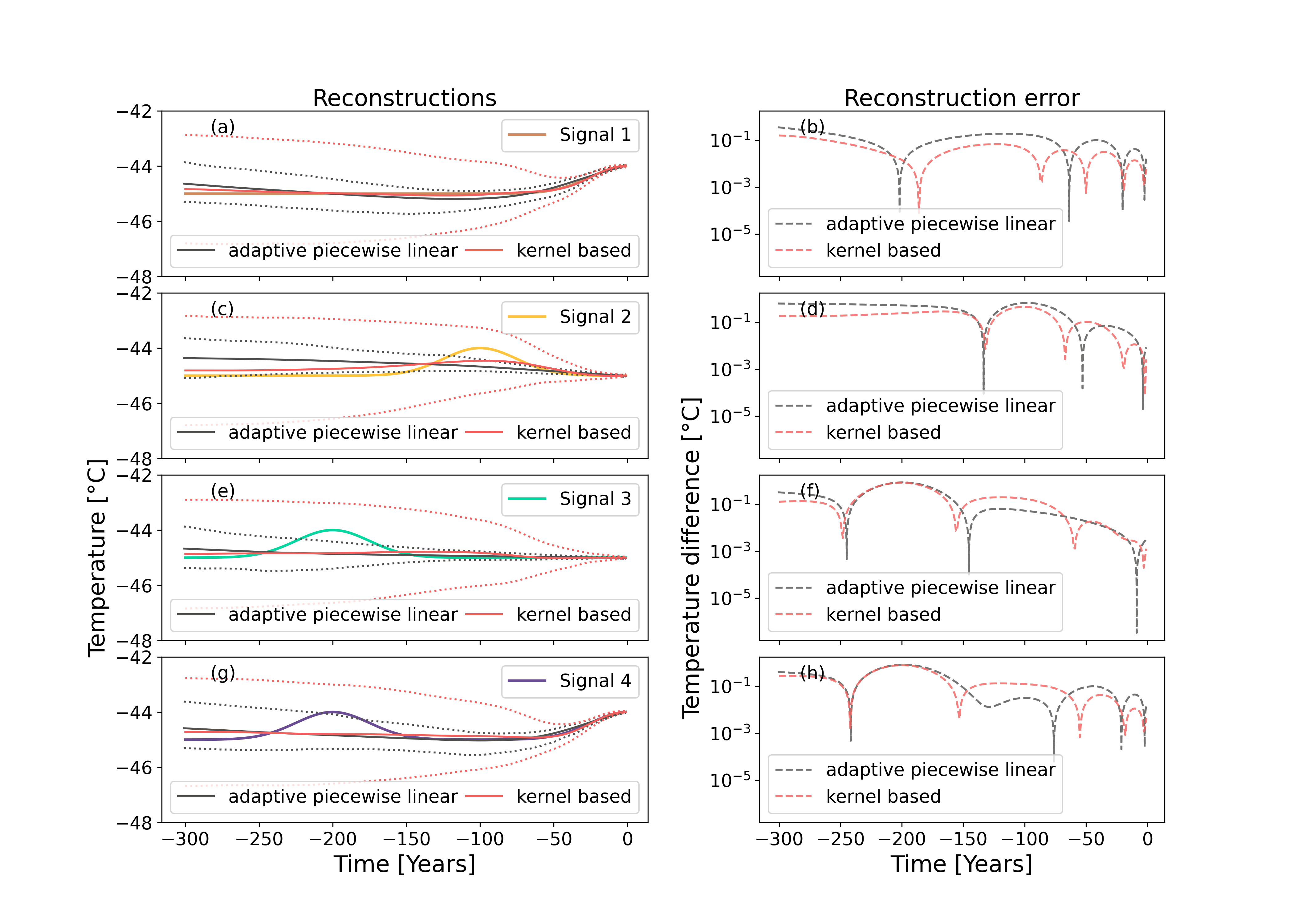}
  \caption{Comparing reconstructions with kernel-based model using emcee vs adaptive piecewise-linear model using RJ-MCMC. Panels (a), (c), (e), and (g) show signal reconstructions obtained using measurement uncertainty with $\sigma_m$ of $30 ~\mathrm{mK}$ for RJMCMC (black), and emcee (red). Solid lines denote the posterior mean, and dotted lines represent the corresponding pointwise credible interval. (b),(d),(f) and (h) shows the difference between posterior mean and the synthetic signal.}
\end{figure} 
The reconstructions are performed using the synthetic measurements shown in Fig.~\ref{fig:fig_art_data} with a measurement uncertainty of $30 ~\mathrm{mK}$. Following \cite{Hopcroft2007inference,muto2011multidecadal}, the RJ-MCMC sampler was run to collect a total of 500,000 sampled surface temperature histories, of which the final 400,000 were used to approximate the posterior distribution.

The posterior mean reconstructions obtained with both approaches are generally comparable (Fig.~\ref{fig:fig_rjmcmc} b,d,f,h).  However, the posterior uncertainty estimated by the RJ-MCMC implementation for adaptive piecewise-linear model differs substantially from that obtained with the kernel-based framework. Under the considered experimental setup, the 95\% pointwise credible intervals of the RJ-MCMC reconstructions frequently do not encompass the target surface temperature history (Fig.~\ref{fig:fig_rjmcmc} a,c,e,g). A comparable underestimation of the posterior spread is visible in the RJ-MCMC reconstructions of \cite{muto2011multidecadal} (their Figure 4.25), suggesting that this
behaviour is not specific to our implementation. We cannot exclude that extensive, signal specific hyperparameter tuning would improve the RJ-MCMC results; the hyperparameters
used here are those recommended by \cite{Hopcroft2007inference, muto2011multidecadal} and are listed in Appendix \ref{rjmcmc_appendix}. 

From a computational perspective, the differences are also substantial. The RJ-MCMC implementation generated 500,000 posterior samples, whereas the ensemble MCMC sampler collected approximately 14.6 million samples (122 walkers with 120,000 samples each) in less than half the time. Experiments with lower assumed measurement uncertainties are computationally more expensive using the reference RJ-MCMC implementation, while the kernel-based framework using affine invariant ensemble MCMC sampler readily accommodates reconstructions with measurement uncertainties of $10 ~\mathrm{mK}$ and $1 ~\mathrm{mK}$ without additional tuning (Fig.~\ref{fig:fig_merr}).

We emphasize that the purpose of this comparison is not to evaluate the adaptive piecewise-linear representation itself, but to compare two Bayesian inversion frameworks for shallow borehole thermometry. Under the considered experimental setup, the kernel-based framework is easy to implement and provides more reliable posterior uncertainty estimates while enabling substantially more efficient representation and exploration of the solution space.

\section{Case study with realistic surrogate climate histories}
\label{sec:case_study}
The synthetic experiments conducted in Section \ref{sec:reconstruction using kbm} benchmarks the performance of the proposed kernel-based model for surface temperature reconstructions under idealized conditions. However, in a real world scenario, the surface temperature evolution consists of substantial short-term climate variability that cannot be fully represented by the kernel-based surface temperature model. The question that arises is how such unresolved variability affects surface temperature reconstructions.

To investigate this, we use surrogate surface temperature time series generated using two components: a long-term warming trends derived from global mean temperature (GMT) multiplied by a polar amplification factor $PA=0,1,2,3$, and realistic local climate variability represented by a power-law spectrum with scaling exponent $\beta=0.6$ \cite{shaju2025sensor}. These surrogate surface temperature time series thus replicates realistic scenario by mimicking both long-term climate trends and natural year-to-year variability.

\begin{figure}
  \centering
  \label{fig:case_study_signals}
  \includegraphics[width=\textwidth]{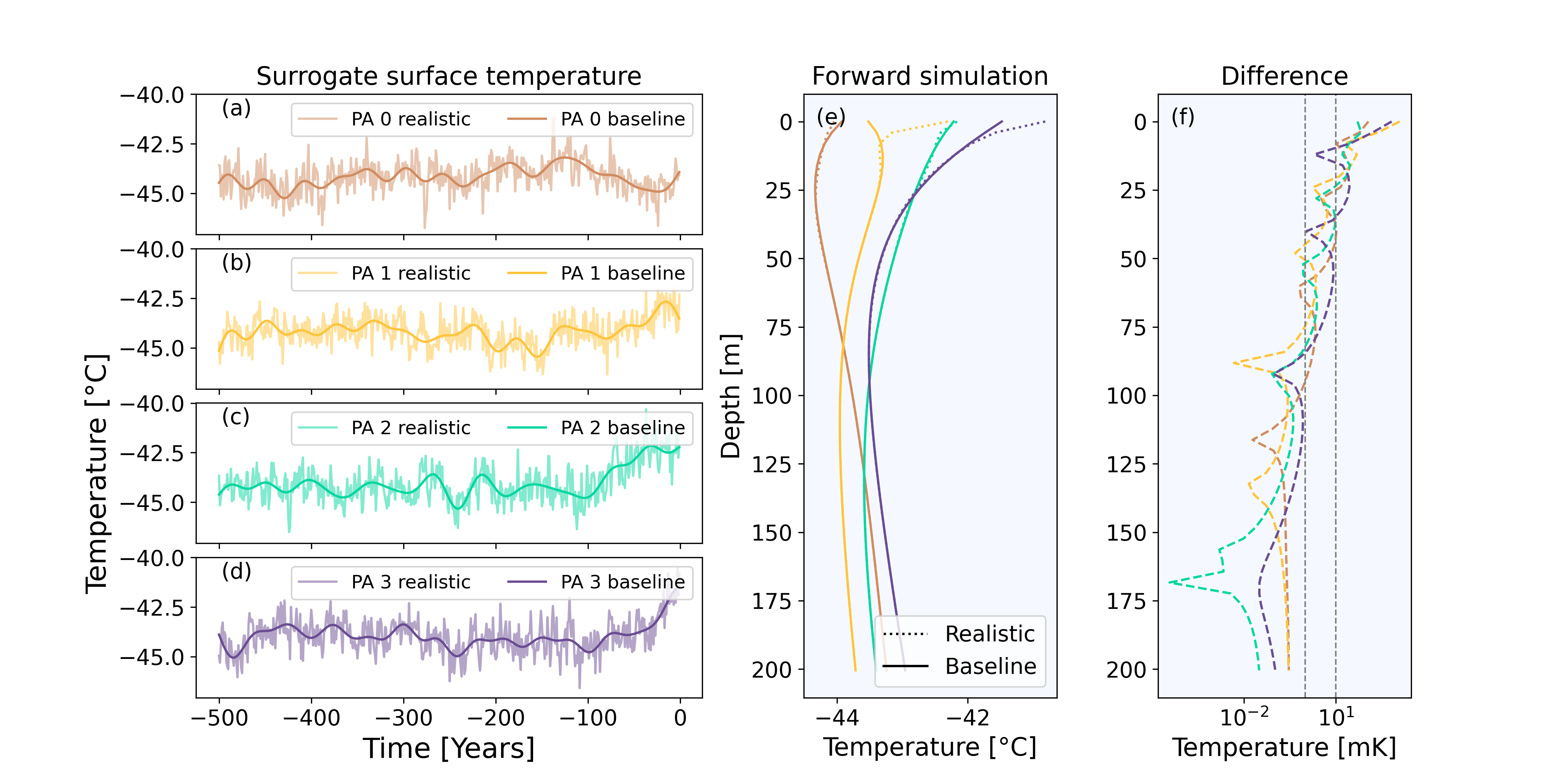}
  \caption{Realistic surrogate surface temperature history and data. (a), (b), (c), and (d) show the realistic synthetic surface temperature realizations (light colored solid curve) and their respective baseline signals (dark colored solid curve). Forward model simulations of realistic (dotted curve) and baseline (solid curve) signals are shown in (e), while (f) shows the difference between the forward model simulations of the realistic signals with respect to their baseline signals.}
\end{figure}

For this study we consider individual realizations of each of the four categories ($PA=0,1,2,3$) of realistic temperature series (Fig.~\ref{fig:case_study_signals}a-d). As mentioned earlier, since the kernel-based surface temperature model cannot precisely represent short-term climate variability, we define a smoothed representation of each surrogate time series, referred to as the baseline signal. The baseline signal corresponds to the best possible approximation of the surrogate time series that can be achieved with the chosen kernel configuration (40 kernels, length-scale 20 years). Our objective is to reconstruct these baseline signals rather than the full surrogate temperature series. 

Forward simulations of the realistic and baseline signals produce slightly different borehole temperature profiles (Fig.~\ref{fig:case_study_signals}e,f). This discrepancy originates from limitations of the surface temperature model itself, and we interpret this discrepancy as surface temperature model approximation uncertainty \cite{Cvetkovi2024}. The differences amount to several $\mathrm{mK}$ at depths of approximately $30–100 ~\mathrm{m}$ and increase to about $1 ~\mathrm{K}$ near the surface (Fig.~\ref{fig:case_study_signals}f), highlighting that unresolved climate variability can exceed the measurement uncertainty considered so far.

We investigate three complementary scenarios: first, reconstructions using ideal measurements generated directly from the baseline signals (Section \ref{sec: baseline_signals_reconstructions}), second, reconstructions using realistic measurements while accounting for approximation uncertainty arising from unresolved climate variability (Section \ref{sec:fig_realistic_signal_reconstructions}), and finally reconstructions using an idealized upper-bound scenario assuming perfect knowledge of the climate variability (Section \ref{sec:fig_known_anomaly_reconstructions}). For all experiments, synthetic measurements are extracted of respective forward simulations at 40 equally spaced depths between 0 and 200 m. The Bayesian inversion is performed using the same kernel configuration throughout this section (40 kernels, length-scale 20 years). We collected $150,000$ samples (per walker), of which the first $25,000$ samples (per walker) were discarded as burn-in. The priors are selected as $\theta_{\mathrm{pom}} \sim \mathcal{U}(-50, -40)$ and $\alpha_i \sim \mathcal{N}(0, 1.2^{2}\mathbf{I})$. A measurement uncertainty of $1 ~\mathrm{mK}$ is assumed for Section \ref{sec: baseline_signals_reconstructions} and Section \ref{sec:fig_known_anomaly_reconstructions}. In Section \ref{sec:fig_realistic_signal_reconstructions}, the approximation uncertainty estimated from an ensemble of surrogate realizations is also considered in the likelihood in-addition to the measurement uncertainty.

\subsection{Idealized benchmark: reconstruction of baseline signals}

\begin{figure}[h]
  \centering
  \label{fig: baseline_signals_reconstructions}  \includegraphics[width=\textwidth]{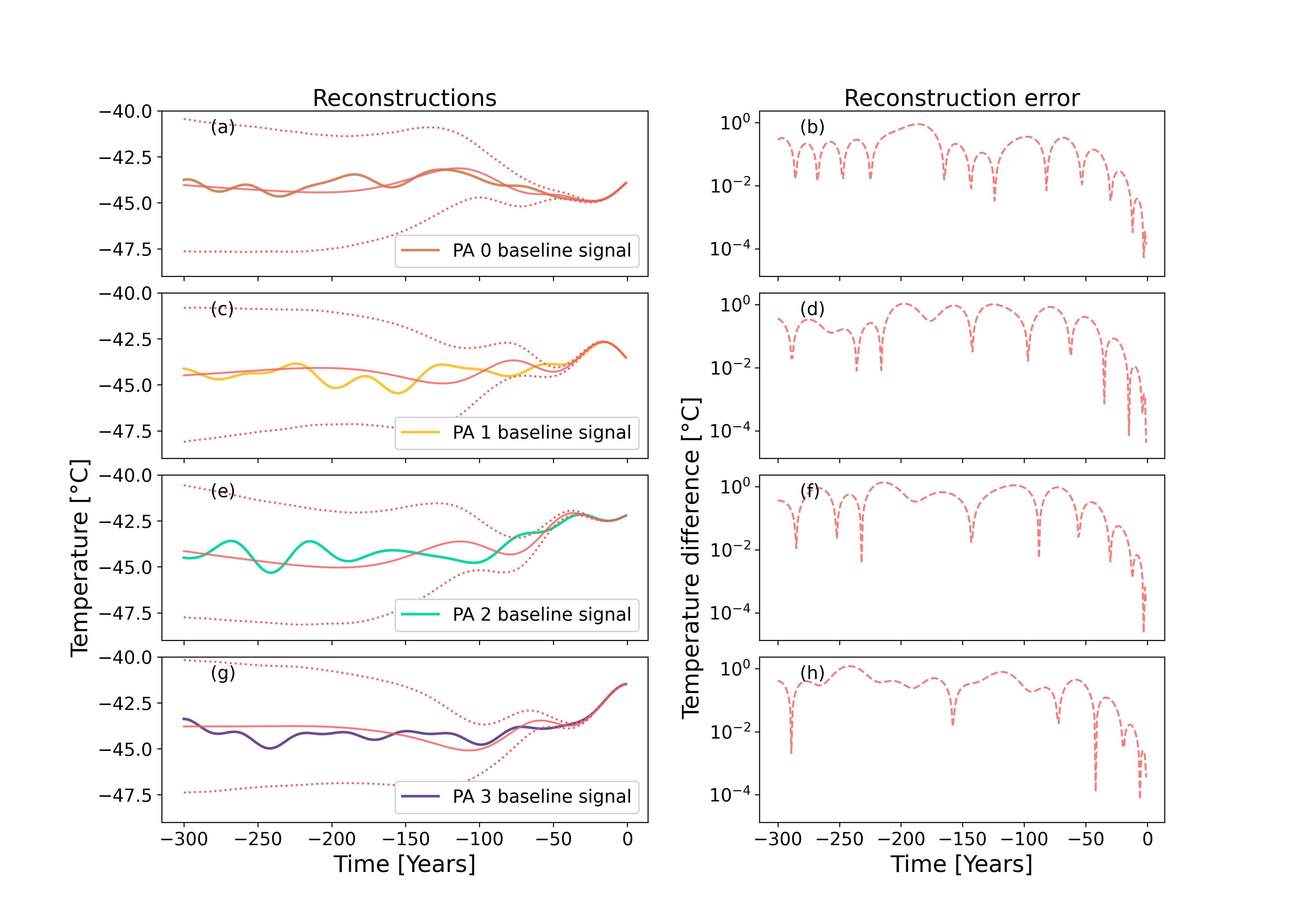}
  \caption{Reconstructions with the most ideal data,  generated using baseline signals. (a), (c), (e), and (g) show the reconstructions of the baseline signals for PA 0, 1, 2, and 3 using  measurement uncertainty with $\sigma_m$ of $1 ~\mathrm{mK}$. The red solid line shows the posterior mean, and the dotted line shows the point wise credible interval. (b), (d), (f), and (h) shows the absolute difference between the posterior mean and the corresponding baseline signal.}
\end{figure} 
\label{sec: baseline_signals_reconstructions} 
The ideal measurements contain information only about the baseline signals and are free from additional discrepancies caused by unresolved climate variability. We perform reconstructions using 40 measurement points and assume a measurement uncertainty with a standard deviation of $1 ~\mathrm{mK}$. Under these idealized conditions, the baseline signals are reconstructed accurately near the present, while the quality of the reconstructions gradually deteriorates further back in time (Fig.~\ref{fig: baseline_signals_reconstructions}a,c,e,g). Reconstruction errors are on the order of $1 ~\mathrm{mK}$ during the most recent 10 years, increase to approximately $10 ~\mathrm{mK}$ over the last 10–25 years, exceed $100 ~\mathrm{mK}$ after about 50 years, and reach around $1 ~\mathrm{K}$ by 100 years for most cases (Fig.~\ref{fig: baseline_signals_reconstructions}b,d,f,h). Beyond this point, the error remains approximately constant. The reconstruction performance is largely independent of the specific surrogate time series considered.

\begin{figure}[htbp]
  \centering
  \label{fig:ensemble reconstructions}
  \includegraphics[width=\textwidth]{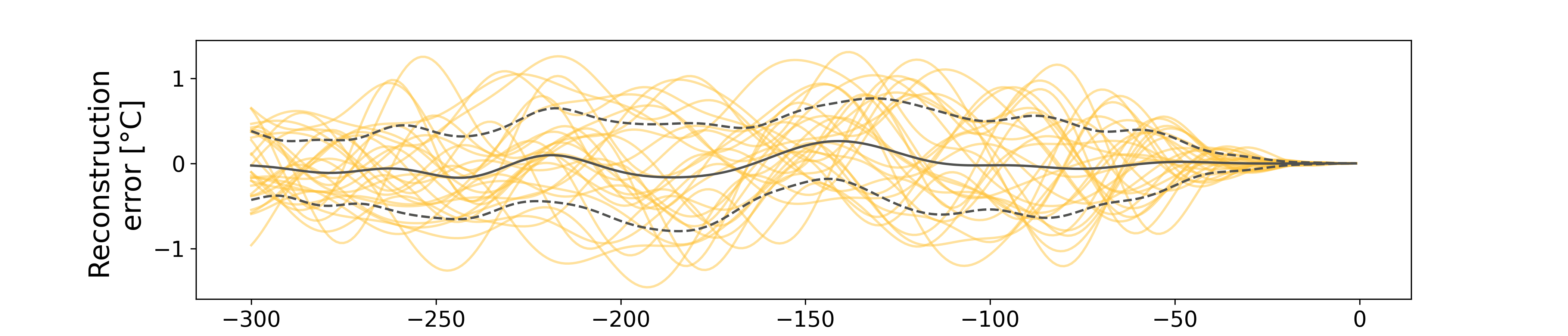}
  \caption{Ensemble of reconstruction errors, obtained from reconstructions using measurements generated from 28 distinct baseline time series of type PA = 1. The black solid line denotes the mean error, while the dashed line denotes the standard deviation of the error at each point in time.}
\end{figure}

To assess potential systematic biases, we additionally performed reconstructions for 28 individual baseline signals of type PA = 1, and analyzed their reconstruction errors. The ensemble of reconstruction errors (Fig.~\ref{fig:ensemble reconstructions}) shows that the mean error remains close to zero throughout the entire time interval, indicating no systematic bias.
The spread of the reconstruction error increases with age, reaching a standard deviation of approximately 0.5\,$^\circ\mathrm{C}$ by 80--100 years, after which it stabilizes.

\subsection{Reconstructions using realistic measurements and approximation uncertainty}
 \label{sec:fig_realistic_signal_reconstructions}

\begin{figure}[htbp]
  \centering
  \label{fig:fig_realistic_signal_reconstructions}
  \includegraphics[width=\textwidth]{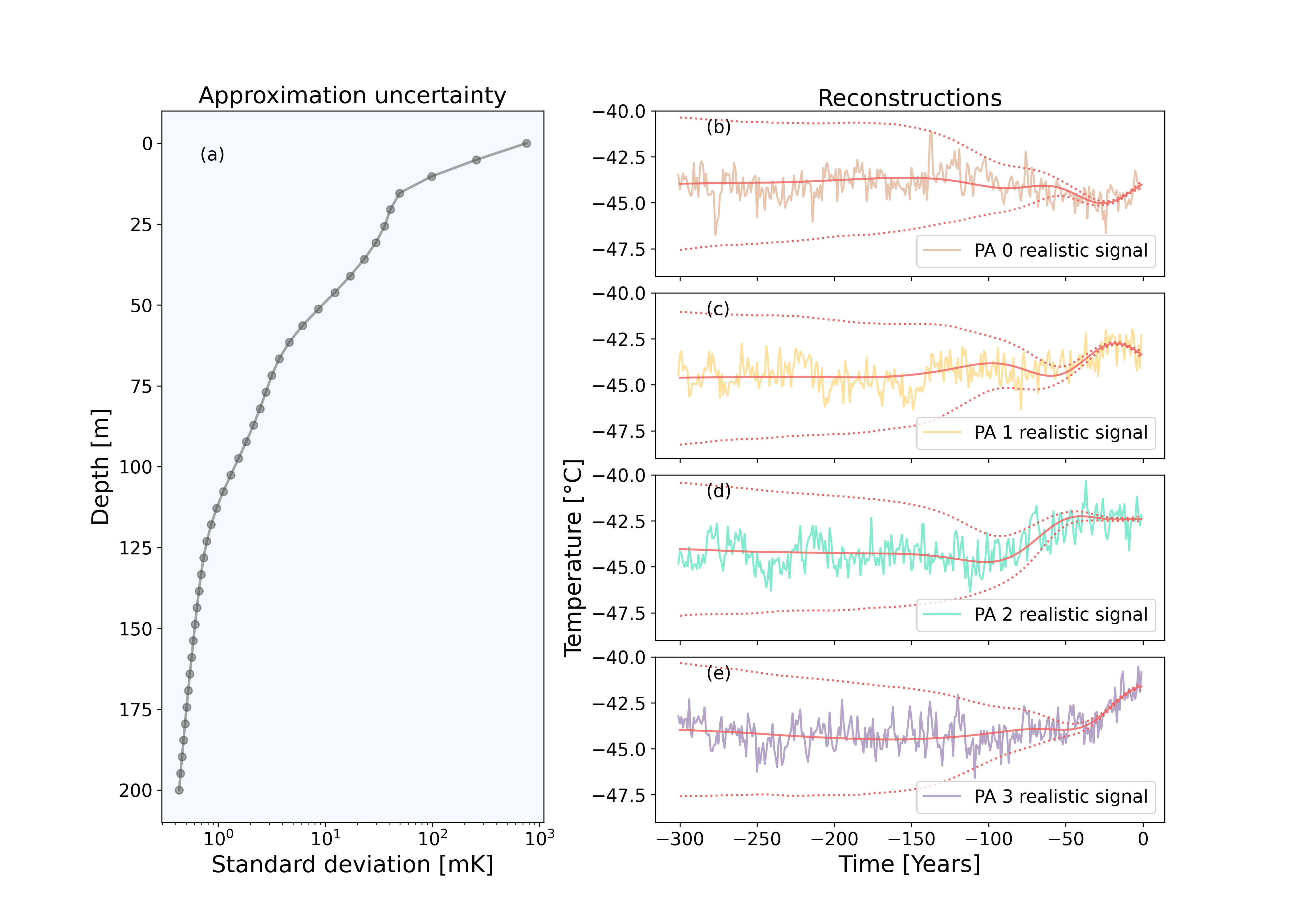}
  \caption{Reconstructions with data generated using realistic signals considering surface temperature model approximation uncertainty. (a) shows depth-wise standard deviations computed to represent the surface temperature model approximation uncertainty. (b), (c), (d), and (e) show the reconstructions of the realistic signals for PA 0, 1, 2, and 3 using measurements from realistic signals considering approximation uncertainty.}
\end{figure} 
 
Realistic borehole measurements contain information from both the long-term temperature trend and unresolved high-frequency climate variability. Since the kernel-based surface temperature model cannot represent this variability, attempting to explain highly resolved measurements may lead to misleading reconstructions. Our objective therefore remains to reconstruct the baseline signals from realistic measurements. We quantify the surface temperature model approximation uncertainty using 1000 realizations of each surrogate category ($PA = 0, 1, 2, 3$). It is quantified as a normal distribution, with the standard deviation at each depth point computed as the root-mean-square error between forward simulations of the realistic and baseline signals. We consider this approximation uncertainty in the likelihood together with the measurement uncertainty. 

Accounting for approximation uncertainty, the posterior mean successfully recovers the overall temperature evolution and captures the main trends approximately over the past 100 years (Fig.~\ref{fig:fig_realistic_signal_reconstructions}b-e), with particularly robust reconstructions over the past 50 years. Note that these reconstructions (Fig.~\ref{fig:fig_realistic_signal_reconstructions}b-e) are performed with only approximation uncertainty, assuming no measurement uncertainty. Compared to the idealized experiments (Fig.~\ref{fig: baseline_signals_reconstructions}), posterior uncertainties are substantially larger because approximation uncertainty dominates over measurement uncertainty. Consequently, reducing measurement uncertainty alone does not necessarily improve reconstructions under realistic conditions.


\subsection{Reconstructions assuming perfect knowledge of climate variability}
  \label{sec:fig_known_anomaly_reconstructions}
\begin{figure}[htbp]
  \centering
  \label{fig:fig_known_anomaly_reconstructions}
  \includegraphics[width=\textwidth]{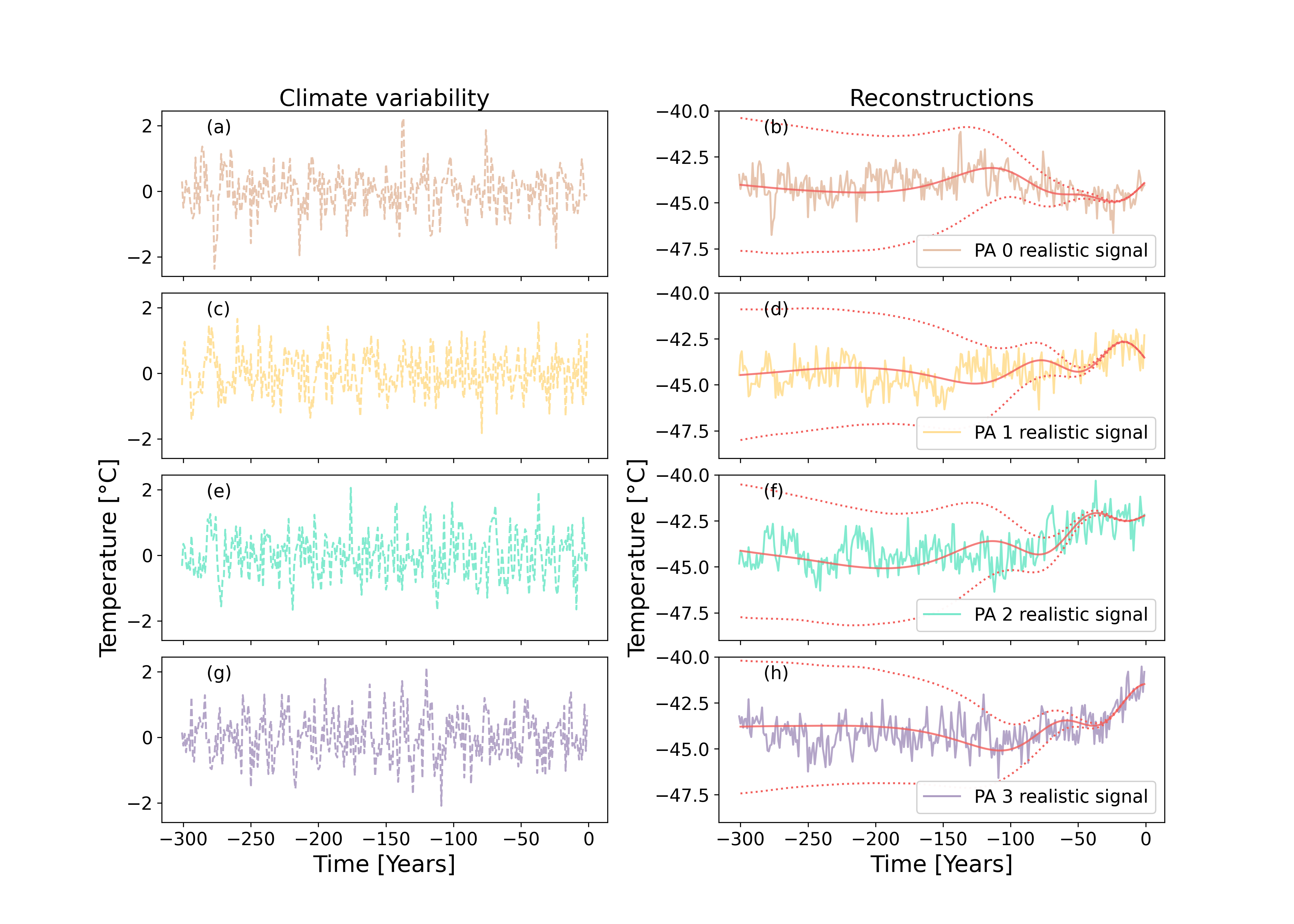}
  \caption{Reconstructions with data generated using realistic  signals and known climate variability. (a), (c), (e), and (g) shows the known climate variability. (b), (d), (f), and (h) show the reconstructions using a measurement uncertainty with a standard deviation of $1 ~\mathrm{mK}$, and including climate variability as a known parameter in the forward model.}
\end{figure} 

To isolate the impact of unresolved climate variability, we perform an idealized experiment in which the exact climate variability component is assumed to be known. The noise representing exact climate variability is extracted from each realistic temperature series by subtracting its respective baseline signal (Fig.~\ref{fig:fig_known_anomaly_reconstructions}a,c,e, and g). This extracted noise is added to every sampled surface temperature, and the resulting signal is supplied to the forward model for the likelihood evaluation. This allows us to assess the upper performance limit that could be achieved with realistic measurements. Under this assumption, reconstructions obtained from realistic measurements with a measurement uncertainty of $1 ~\mathrm{mK}$ closely resemble those obtained using the idealized baseline measurements (Fig.~\ref{fig:fig_known_anomaly_reconstructions}b,d,f,h and Fig.~\ref{fig: baseline_signals_reconstructions}a–d). This demonstrates that the degradation in reconstruction observed in Section \ref{sec:fig_realistic_signal_reconstructions} is caused by unknown unresolved climate variability rather than by limitations of the inversion framework itself. This experiment is intended as a diagnostic upper-bound estimate and does not represent a realistic application, since the exact realization of climate variability is generally unknown.

\section{Conclusions}
\label{sec:conclusions}

We presented a kernel-based surface temperature model for reconstructing past surface temperature histories from shallow ice borehole temperature measurements within a Bayesian framework. The proposed surface temperature model enables the use of parallel ensemble Markov chain Monte Carlo sampler (emcee), providing efficient exploration of the solution space.

Using synthetic experiments, we systematically investigated the influence of kernel configuration, measurement density, measurement uncertainty, and temporal smearing on reconstruction performance. We found that reconstruction quality is largely insensitive to the number of kernels once the kernel basis is sufficiently dense, and that increasing the number of measurement points for fixed depth provides only marginal improvement. In contrast, reducing measurement uncertainty substantially improves reconstruction quality under idealized conditions.

The comparison with the reference adaptive piecewise-linear surface temperature model using RJ-MCMC sampling highlights the practical advantages of the proposed inversion framework. Under the considered experimental setup, the kernel-based approach provides more reliable posterior uncertainty estimates while enabling substantially more efficient solution space exploration.

We further evaluated the proposed framework using realistic surrogate surface temperature histories containing stochastic climate variability. These experiments show that unresolved high-frequency variability introduces model approximation uncertainty because it cannot be represented by the chosen surface temperature model. When this approximation uncertainty exceeds instrumental measurement uncertainty, it becomes the dominant source of uncertainty in the inversion. Consequently, further improvements in measurement precision alone do not necessarily translate into more accurate climate reconstructions.

We find that reconstructions from measurements unaffected by short-term variability resemble those obtained from measurements affected by short-term variability when this variability is perfectly known, confirming the performance of the kernel-based inversion framework. This highlights the importance of quantifying uncertainty arising from unresolved short-term climate variability, which cannot be fully captured by the surface temperature models.

Our results therefore demonstrate that reliable Bayesian inversion of ice borehole temperature measurements requires accounting for both measurement uncertainty and model approximation uncertainty. The proposed kernel-based framework provides an efficient basis for such reconstructions while offering a practical route towards uncertainty-aware reconstructions of past surface temperature histories from shallow ice boreholes.

\appendix

\section{Forward model setup} 
\label{fwd_appendix}
The calculations of the vertical velocity profile $w$ and the thermal diffusivity profile $k$, are provided herein. The thermal diffusivity is calculated as 
\begin{equation}\label{eq2}
k=\frac{K}{\rho c},
\end{equation}
where $c$ is the specific heat capacity, $K$ is the thermal conductivity, and $\rho$ is the density of the firn/ice.  

The density profile is simulated using the Herron- Langway model \cite{Herron_1980, Arthern_2010}, which requires surface snow density $(\rho_{s})$ in addition to $\theta_{m}$, $w_{s}$ and $H$. The unit of density is $\mathrm{kg\ m^{-3}}$ and it is one of the key inputs to the heat transfer model to calculate heat capacity $c$ and thermal conductivity $K$. The equations for their calculations are listed below \cite{Cuffey_Paterson_2010, Muto2011recent, Orsi2012little}.

\subsection{
Specific heat capacity ($c$)}
Specific heat capacity of ice ($c_{ice}$) is given by the following \cite{Paterson_1994},
\begin{equation}\label{eq_ci}
c_{ice}= 152.5 + 7.122 T\,,
\end{equation}
where $T$ is the temperature in Kelvin. 
Specific heat capacity of firn ($c_{firn}$) is calculated from the percentage of ice and air in firn which is given by,
\begin{equation}\label{eq_cfirn}
c_{firn} = c_{ice}\frac{\rho}{\rho_{ice}} + c_{a}(1-\frac{\rho}{\rho_{ice}})\,,
\end{equation}
where $\rho_{ice}$ is the density of ice (917 $\mathrm{kg\ m^{-3}}$) and $c_{a}$ (1005 $\mathrm{J\ kg^{-1}\ K^{-1}}$) is the specific heat capacity of dry air. 

\subsection{Thermal conductivity ($K$)}
The temperature-dependent thermal conductivity of ice ($K_{ice}$) in $\mathrm{W\ m^{-1}\ K^{-1}}$ is given by
\begin{equation}\label{eq_Ki}
K_{ice} = 2.22(1-0.0067 T)\,,
\end{equation}
where T is temperature in \textdegree \,C. 
Thermal conductivity of firn ($K_{firn}$) is given by 
\begin{equation}\label{eq_Kfirn}
K_{firn} = K_{ice}\Biggl(\frac{\rho}{\rho_{ice}}\Biggr)  ^{\alpha^{\prime}-\beta^{\prime}\Biggl(\frac{\rho}{\rho_{ice}}\Biggr)}\,,
\end{equation}
where $\alpha^{\prime}$ and $\beta^{\prime}$ are site specific coefficients. We used $\alpha^{\prime}=2.4634$ and $\beta^{\prime}=0$ for the experiments conducted in this study which was used by \cite{muto2011multidecadal} for the site NUS07-2.

\subsection{Vertical velocity ($w$)}
 As in \cite{muto2011multidecadal}, the velocity profile is computed using the equation from \cite{Goujon_2003}, which is based on the ice velocity model by \cite{Lliboutry_1979}. Using a relative vertical coordinate $\zeta = z/H$, where H is the ice sheet thickness, $w(\zeta)$ is given as
\begin{equation}\label{eq_Kfirn}
w(\zeta) = \frac{\rho}{\rho_{ice}} \Biggl[w_{s}-(w_{s}-w_{b}) \Biggl(\frac{m+2}{m+1}\zeta\Biggr)\Biggl(1-\frac{\zeta^{m+1}}{m+2}\Biggr) \Biggr]\,.
\end{equation}
The constants $w_{s}$ and $w_{b}$ are the vertical velocity at the surface and at the base of the ice sheet, respectively. $w_{s}$
 is assumed to be equal to the accumulation rate and $w_{b}$ is the basal melting rate.
$m$ is the shape parameter of the vertical velocity profile and is set to $11$ \cite{shaju2025sensor}.

\subsection{Initial condition profiles for inverse model}
The initial condition during the inversion process, is selected based on the initial value of the sampled kernel-based surface temperature time series from a precomputed set of initial condition profiles spanning all temperature values between $-40.0$ to $-50.0 ^\circ \mathrm{C}$ at a resolution of up to $1 ~\mathrm{mK}$. For the experiments with the resolution of measurement uncertainty $<1 ~\mathrm{mK}$, we compute the intermediate initial condition profile. This is done by taking the residuals of consecutive $1 ~\mathrm{mK}$ profiles, multiplying by the factor as required, and adding it to the minimum of $1 ~\mathrm{mK}$ profile. The initial condition for adaptive piecewise-linear model is selected based on the sampled $\theta_{pom}$.

\section{RJMCMC setup}
\label{rjmcmc_appendix}

\subsection{Priors}
\label{rjmcmc_prior}
The prior distribution on the time locations is based on order statistics which is given by:
\begin{equation}
p({t}) =
\frac{k!}{D^{k}}
\, \mathbb{I}\!\left( 0 < t_1 < t_2 < \cdots < t_k < D \right),
\label{eq:order_stat_prior}
\end{equation}
where $D$ is the length of the time domain ($t_{max} - t_{min}$) and $\mathbb{I}$ is the order statistics uniform distribution \cite{Hopcroft2007inference}. Here, $k$ denotes the number of points in a piecewise-linear model, we take dimensionality prior $p(k)$ as a uniform distribution using minimum and maximum value of $k$ \cite{Hopcroft2007inference}. 

\subsection{Proposal density} 
\label{rjmcmc_proposl}
 At each iteration of the RJ-MCMC algorithm, one of the below specified modification is chosen at random, giving probability of $1/5$ for choosing a move. The probability is $1/3$ when the number of time points equals the predefined maximum value of $k$, and $1/2$ when it equals the predefined minimum value.
\begin{enumerate}
    \item \textbf{Modifying $\theta(t_j)$ at random position $j$:}
    \begin{equation}
        \theta(t_j)' = \theta(t_j)+ u_2 \, \sigma_m^{T} \, \exp\!\left(\frac{k-j}{k}\right),
        \label{eq:temperature_update}
    \end{equation}
    where $u_2 = \sim \mathcal{N}(0, 1^2)$, $\sigma_m^{T} = 0.1$.
    \item \textbf{Modifying $t_j$ at random position $j$:}
    \begin{equation}
        t'_j = t_j + \left[ t^- - t^+ \right] \, u_1 \, \sigma_m^{t} \,
        \exp\left(\frac{k-j}{k}\right)
        \label{eq:t_update}
    \end{equation}
where $u_1 = \sim \mathcal{N}(0, 1^2)$, $\sigma_m^{t} = 0.05$. We use $\frac{k-j}{k}$ instead of $\frac{j}{k}$ as we implement $t_0$ at the starting indices of the arrays, in contrast to \cite{Hopcroft2007inference}.
    
\item \textbf{Adding a new point $(t_j,\theta(t_j))$:}
    \begin{equation}
        t^{*} = t^{-} + \sigma_b^{t} \, u_1 \left( t^{+} - t^{-} \right),
        \label{eq:t_birth}
    \end{equation}
    and, 
    \begin{equation}
        \theta(t^{*}) =
        \frac{\theta(t^{+}) - \theta(t^{-})}{t^{+} - t^{-}}
        \left( t^{*} - t^{-} \right)
        + u_2 \sigma_b^{T} + \theta(t^{-}),
        \label{eq:T_birth}
    \end{equation}
    where, $t^+,t^-$ represent selected point $t_j$ and previous time point $t_{j-1}$, $u_1 = \sim \mathcal{U}(0, 1)$, $u_1 = \sim \mathcal{N}(0, 1^2)$ and $\sigma_b^{t} = 1$. We used $\sigma ^b_T =10^{-3}$ instead of $2.5 \times 10^{-4}$ (as in \cite{Muto2011recent}) as it was slightly faster at identifying the sample time series.

    \item \textbf{Deleting an existing point:}
    A random point $(t_j,\theta(t_j))$ is selected and removed, after which the adjacent points are connected.
    \item \textbf{Modifying $\theta_{pom}$:}
    New value is randomly chosen from uniform distribution.
    \end{enumerate}

\subsection{Calculating acceptance probability}

RJ-MCMC acceptance is computed as the minimum of 1 and the product of the prior ratio, likelihood ratio, proposal ratio, and Jacobian correction term (i.e., calculation of $q$ in Algorithm \ref{alg_rjmcmc}). 


\begin{enumerate}
    \item \textbf{Modifying temperature:}

    \begin{equation}
\label{eq:temperature_ac_ratio}
q= min\left( 1 ,\frac{p(\mathcal{T^*})}{p(\mathcal{T})}\frac{p(\vec{y}\mid S^*)}{p(\vec{y}\mid S(i))} \right),
\end{equation}
where $\mathcal{T^*}$ and $\mathcal{T}$ represent set temperature values before and after updation of $\theta(t_j)$.

    \item \textbf{Modifying time point:}
\begin{equation}
\label{eq:time_point_ac_ratio}
q= min\left( 1 ,\frac{(t^{+}-t'^{*})(t'^{*}-t^{-})}{(t^{+}-t^{*})(t^{*}-t^{-})}\frac{p(\vec{y}\mid S^*)}{p(\vec{y}\mid S(i))} \right)
\end{equation}

    \item \textbf{Birth move-- Adding a new point $(t_j,\theta(t_j))$:}
    Prior ratio needs to include prior on Temperature, the number of time steps and the spacing using the order statistics (Equation (\ref{eq:order_stat_prior})). The combined prior ratio is,
    \begin{equation}
    \label{eq:combi_prior_ratio}
    \frac{p(S^*)}{p(S(i))}=\frac{p(\mathcal{T^*})}{p(\mathcal{T})}\frac{p(k+1)}{p(k)}\frac{(k+1)}{D}\frac{(t^{+}-t'^{*})(t'^{*}-t^{-})}{(t^{+}-t^{*})(t^{*}-t^{-})},
    \end{equation}
the proposal ratio is,
    \begin{equation}
\frac{Q(S(i);S^*)}{Q(S^*;S(i))}=\frac{d_{k+1}1/(k+1)}{b_{k}1/D}=\frac{D d_{k+1}}{(k+1)b_{k}}, 
\end{equation}
where, $d_{k+1} =b_k=\frac{1}{5}$. When $k = k_{min}$ only two options are allowed: birth or perturb a temperature, therefore, $d_{k_{min}+1} = \frac{1}{5}, b_{k_{min}}=\frac{1}{2}$.   

The Jacobian is,
\begin{align}
|J| = \left|
\frac{\partial (\theta(t^{*}),t^{*})}{\partial (u_{1},u_{2})}
\right| =
\begin{vmatrix}
(\theta(t^{+})-\theta(t^{-}))\sigma_{t} & \sigma_{T} \\
\sigma_{t}(t^{+}-t^{-}) & 0
\end{vmatrix} 
= -\sigma_{T}\sigma_{t}(t^{+}-t^{-}).
\label{jacobian}
\end{align}

We calculate acceptance as the product of the prior ratio, likelihood ratio, proposal ratio, and Jacobian.

\item \textbf{Death move-- Deleting an existing:}
The acceptance term for a death move has the same form as that for a birth move, with the current model of dimension $k$ proposing a move to dimension $k-1$ and the ratios are inverted, with the appropriate relabelling of dimension-dependent terms \cite{Hopcroft2007inference, green1995reversible}. When $k = k_{max}$ only three options are allowed: death or perturb a temperature or time, therefore, $d_{k_{max}} = \frac{1}{3}, b_{k_{max}-1}=\frac{1}{5}$. 

\end{enumerate}

\section*{Code and data availability}
The code and data for this study are available in our GitHub repository:~\url{https://github.com/SM4DA/IceBT_Bayesian_Inversion_ST}.

\section*{Declarations}
ChatGPT (OpenAI) was used to improve the readability of the manuscript (e.g., sentence structure and grammar) and to assist in generating portions of the Matplotlib visualization code (e.g., axes and legend formatting). All text were reviewed and code were adapted, and validated by the authors.

\section*{Acknowledgments}

This project was supported by the Helmholtz School for Marine Data Science (MarDATA) through grant HIDSS-0005.

We gratefully acknowledge the advice and support provided by Hugo Beltrami and Adalbert Wilhelm, as well as the support of the Interdisciplinary Center for Machine Learning and Data Analytics (IZMD) at the University of Wuppertal. We also thank Nora Hirsch and Andrew Dolman for their fruitful discussions.

\bibliographystyle{siamplain}
\bibliography{references}
\end{document}